\documentclass[10pt, a4paper, preprint]{article}
\pdfoutput=1
\usepackage{anysize}
\marginsize{2cm}{2cm}{1cm}{1cm}

\usepackage{amsmath}
\usepackage{amssymb}
\usepackage{mathtools}
\usepackage{amsthm}

\usepackage{wasysym}
\usepackage{stix}
\usepackage{authblk}
\usepackage[bottom]{footmisc}
\usepackage{graphicx}
\usepackage[font={small,it}]{caption}
\usepackage{bm}
\usepackage{units}
\usepackage{centernot}
\usepackage{hyperref}

\usepackage{caption}
\usepackage{subcaption}
\usepackage{graphicx}
\usepackage{framed}

\usepackage{tikz}
\usetikzlibrary{positioning}
\usetikzlibrary{matrix}
\usetikzlibrary{fit}
\usetikzlibrary{backgrounds}
\usetikzlibrary{shapes}
\usetikzlibrary{decorations.markings}
\usetikzlibrary{tikzmark}
\usetikzlibrary{decorations.markings}
\usetikzlibrary{fadings}
\usepackage{tikz-3dplot}
\usetikzlibrary{patterns}
\usetikzlibrary{plotmarks}

\makeatletter
\long\def\pgfshapeaddanchor#1#2{%
{%
  \def\pgf@sm@shape@name{#1}%
  \let\anchor=\pgf@sh@anchor%
  #2}%
}
\pgfshapeaddanchor{rectangle}{%
  \anchor{west north west}{%
    \pgf@process{\northeast}%
    \pgf@ya=1.5\pgf@y%
    \pgf@process{\southwest}%
    \pgf@y.5\pgf@y%
    \advance\pgf@y by \pgf@ya%
    \pgf@y=.5\pgf@y%
  }%
  \anchor{north north west}{%
    \pgf@process{\southwest}%
    \pgf@xa=1.5\pgf@x%
    \pgf@process{\northeast}%
    \pgf@x.5\pgf@x%
    \advance\pgf@x by \pgf@xa%
    \pgf@x=.5\pgf@x%
  }%
  \anchor{west south  west}{%
    \pgf@process{\northeast}%
    \pgf@ya=.5\pgf@y%
    \pgf@process{\southwest}%
    \pgf@y=1.5\pgf@y%
    \advance\pgf@y by \pgf@ya%
    \pgf@y=.5\pgf@y%
  }%
  \anchor{south south  west}{%
    \pgf@process{\northeast}%
    \pgf@xa=.5\pgf@x%
    \pgf@process{\southwest}%
    \pgf@x=1.5\pgf@x%
    \advance\pgf@x by \pgf@xa%
    \pgf@x=.5\pgf@x%
  }%
  \anchor{south south east}{%
    \pgf@process{\northeast}%
    \pgf@xa=1.5\pgf@x%
    \pgf@process{\southwest}%
    \pgf@x=1.5\pgf@x%
    \advance\pgf@x by \pgf@xa%
    \pgf@x=.5\pgf@x%
  }%
  \anchor{east south  east}{%
    \pgf@process{\northeast}%
    \pgf@ya=.5\pgf@y%
    \pgf@process{\southeast}%
    \pgf@y=1.5\pgf@y%
    \advance\pgf@y by \pgf@ya%
    \pgf@y=.5\pgf@y%
  }%
}
\makeatother

\makeatletter
\newcommand{\subalign}[1]{%
  \vcenter{%
    \Let@ \restore@math@cr \default@tag
    \baselineskip\fontdimen10 \scriptfont\tw@
    \advance\baselineskip\fontdimen12 \scriptfont\tw@
    \lineskip\thr@@\fontdimen8 \scriptfont\thr@@
    \lineskiplimit\lineskip
    \ialign{\hfil$\m@th\scriptstyle##$&$\m@th\scriptstyle{}##$\hfil\crcr
      #1\crcr
    }%
  }%
}
\makeatother

\usepackage[parfill]{parskip}

\newtheorem{theorem}{Theorem}

\newtheorem{corollary}{Corollary}[theorem]

\theoremstyle{definition}

\tikzfading[name=myfade,
top color=transparent!100, bottom color=transparent!0]

\DeclareMathOperator*{\argmin}{arg\,min}

\begin{document}

\title{On shared and multiple information}
\date{}
\author[1]{Cesare Magri\thanks{You can contact me at: \href{mailto:cesare.magri@shared-information.org}{cesare.magri@shared-information.org}. Tools for computing the quantities defined in this article will be made available on \url{www.shared-information.org}.}}
\maketitle

\section{Introduction}

The goal of this work is to address three outstanding problems in information theory.
\emph{Problem one} is the definition of a non-negative decomposition of the information conveyed by two or more sources about a target variable into the specific contribution of each possible combination of the sources \cite{williams_nonnegative_2010}.
\emph{Problem two} is the definition of a measure of information shared by several sources about the target variable \cite{williams_nonnegative_2010}.
\emph{Problem three} is the definition of a measure of multiple information, that is, the extension of mutual information to more than two variables \cite{yeung_first_2002, timme_synergy_2014}.

We assume that the reader is familiar with Hu and Yeung's set-theoretic structure of Shannon’s information theory \cite{yeung_first_2002, ting_amount_1962} and with the partial information decomposition and partial information diagrams of Williams and Beer \cite{williams_nonnegative_2010}. Here we briefly summarize the main concepts of these theories.

\begin{figure}[t!]
\begin{framed}
\begin{subfigure}[t]{0.4\textwidth}
\centering
	\subcaption[]{\centering Information diagram for two variables}
	\centering
\begin{tikzpicture}
	\def\firstcircle{ (0,0) circle (1)}
	\def\secondcircle{(1,0) circle (1)}
	\def\thirdcircle{ (0.5,-1) circle (1)}

	\node[draw, circle, minimum width = 2 cm, line width = 0.8pt] (I1) at (0,0) {};
	\node[draw, circle, minimum width = 2 cm, line width = 0.8pt] (I2) at (1,0) {};

	\draw[] (120:1.3) -- (120:1);
	\node[anchor = south] at (120:1.3) {\scriptsize{$H(X_1)$}};

	\draw[] (1,0) +(60:1.3) -- +(60:1);
	\node[anchor = south] at ($ (1,0) +(60:1.3) $) {\scriptsize{$H(X_2)$}};
	

	\draw[] (0.5,0) +(-15:0) -- +(-15:1.8);
	\node[anchor = west] at ($ (0.5,0) +(-15:1.8) $) {\scriptsize{$I(X_1;X_2)$}};
	\node[anchor = east, color=white] at ($ (0.5,0) +(195:1.8) $) {\scriptsize{$I(X_1;X_2)$}};
		
	\end{tikzpicture}
\end{subfigure}
\hfill
\begin{subfigure}[t]{0.6\textwidth}
\centering
	\subcaption[]{\centering Information diagram for three variables}
	\centering
\begin{tikzpicture}
	\def\firstcircle{ (0,0) circle (1)}
	\def\secondcircle{(1,0) circle (1)}
	\def\thirdcircle{ (0.5,-1) circle (1)}
%
%
%
%

	\node[draw, circle, minimum width = 2 cm, line width = 0.8pt] (I1) at (0,0) {};
	\node[draw, circle, minimum width = 2 cm, line width = 0.8pt] (I2) at (1,0) {};
	\node[draw, circle, minimum width = 2 cm, line width = 0.8pt] (I3) at (0.5,-1) {};
	
	\draw[] (120:1.3) -- (120:1);
	\node[anchor = south] at (120:1.3) {\scriptsize{$H(X_1)$}};

	\draw[] (1,0) +(60:1.3) -- +(60:1);
	\node[anchor = south] at ($ (1,0) +(60:1.3) $) {\scriptsize{$H(X_2)$}};

	\draw[] (0.5,-1) +(270:1.3) -- +(270:1);
	\node[anchor = north] at ($(0.5,-1) +(270:1.3)$) {\scriptsize{$H(X_3)$}};
	
	\draw[] (0.5,0) +(0,0.3) -- +(-15:1.8);
	\draw[] (0.5,0) +(0,-0.3) -- +(-15:1.8);
	\node[anchor = west] at ($(0.5,0)+(-15:1.8)$) {\scriptsize{$I(X_1;X_2)$}};
	\node[anchor = east, color=white] at ($(0.5,0)+(195:1.8)$) {\scriptsize{$I(X_1;X_2)$}};
	
	\draw[] (0.5,-1) +(0,0.5) -- +(-15:1.3);
	\node[anchor = west] at ($(0.5,-1)+(-15:1.3)$) {\scriptsize{$I(X_1;X_2;X_3)$}};
		
	\end{tikzpicture}
\end{subfigure}

\vspace*{0.3cm}

\begin{subfigure}[t]{0.4\textwidth}
\centering
	\subcaption[short for lof]{Partial information diagram for two sources}
	\centering
\begin{tikzpicture}
		
		\node[draw, circle, minimum width = 2 cm, line width = 0.8pt] (I1) at (0,0) {};
		\node[draw, circle, minimum width = 2 cm, line width = 0.8pt] (I2) at (1,0) {};
		\node[draw, ellipse, minimum width = 4.5 cm, minimum height = 2.8 cm, line width = 0.8pt] (I12) at (0.5,0) {};
		
		
		\node[anchor = south, color=white] at (0.5, 2.44) {\scriptsize{$I(X_1,X_2;Y)$}};
		
		\draw[] (0.5, 1.7) -- (I12.north);
		\node[anchor = south] at (0.5,1.7) {\scriptsize{$I(X_1,X_2;Y)$}};
		
		
		\draw[] (0,0) +(120:1.7) -- +(120:1);
		\node[anchor = south] at ($ (0,0) +(120:1.7) $) {\scriptsize{$I(X_1;Y)$}};
		
		\draw[] (1,0) +(60:1.7) -- +(60:1);
		\node[anchor = south] at ($ (1,0) +(60:1.7) $) {\scriptsize{$I(X_2;Y)$}};
		
		
		
	\end{tikzpicture}
\end{subfigure}
\hfill
\begin{subfigure}[t]{0.6\textwidth}
	\centering
	\subcaption[short for lof]{Partial information diagram for three sources}
	\begin{tikzpicture}	[]

		\node[draw, circle, minimum width = 2 cm, , line width = 0.8pt] (H1) at (0,0) {};
		\node[draw, circle, minimum width = 2 cm, , line width = 0.8pt] (H2) at (1,0) {};
		\node[draw, circle, minimum width = 2 cm, , line width = 0.8pt] (H3) at (0.5,-1) {};
		
		\node[draw, ellipse, minimum width = 4.5 cm, minimum height = 2.8 cm, line width = 0.8pt] (I12) at (barycentric cs:H1=0.5,H2=0.5 ) {};
		\node[draw, ellipse, minimum width = 4.5 cm, minimum height = 2.8 cm, line width = 0.8pt, rotate = 60] (I23) at (barycentric cs:H2=0.5,H3=0.5 ) {};
		\node[draw, ellipse, minimum width = 4.5 cm, minimum height = 2.8 cm, line width = 0.8pt, rotate = -60] (I13) at (barycentric cs:H1=0.5,H3=0.5 ) {};
		
		\node[draw, circle, minimum width = 5 cm, , line width = 0.8pt] (I123) at (0.5, -0.36) {};
		
		\draw[] (0.5, -0.36) +(250:2.5) -- +(250:2.8);
		\node[anchor = north] at ($ (0.5, -0.36) +(250:2.8) $) {\scriptsize{$I(X_1,X_2,X_3;Y)$}};

		\draw[] (0.5, 2.44) -- (I12.north);
		\node[anchor = south] at (0.5, 2.44) {\scriptsize{$I(X_1,X_2;Y)$}};
		
		\draw[] (0,0) +(120:2.2) -- +(120:1);
		\node[anchor = south] at ($ (0,0) +(120:2.2) $) {\scriptsize{$I(X_1;Y)$}};
		
		\draw[] (1,0) +(60:2.2) -- +(60:1);
		\node[anchor = south] at ($ (1,0) +(60:2.2) $) {\scriptsize{$I(X_2;Y)$}};
		
		\draw[] (0.5,-1) +(-60:2.3) -- +(-60:1);
		\node[anchor = north] at ($ (0.5,-1) +(-60:2.3) $) {\scriptsize{$I(X_3;Y)$}};
		
		\draw[] (barycentric cs:H1=0.5,H3=0.5) +(208:1.4) -- +(208:2.5);
		\node[anchor = east, color=white] at ($(barycentric cs:H1=0.5,H3=0.5) +(208:2.5)$) {\scriptsize{$I(X_2,X_3;Y)$}};
		\node[anchor = east] at ($(barycentric cs:H1=0.5,H3=0.5) +(208:2.5)$) {\scriptsize{$I(X_1,X_3;Y)$}};
		
		\draw[] (barycentric cs:H2=0.5,H3=0.5) +(-28:1.4) -- +(-28:2.5);
		\node[anchor = west, color=white] at ($(barycentric cs:H2=0.5,H3=0.5 ) +(-28:2.5)$) {\scriptsize{$I(X_1,X_3;Y)$}};
		\node[anchor = west] at ($(barycentric cs:H2=0.5,H3=0.5 ) +(-28:2.5)$) {\scriptsize{$I(X_2, X_3;Y)$}};
		
%
		
%
%
			
	\end{tikzpicture}
\end{subfigure}
		
\end{framed}
\caption{Examples of information diagrams and partial information diagrams. In information diagrams each variable $X_n$ is assigned a region of size $H(X_n)$. Shannon's mutual information is visualized as the intersection of two regions of size $H(X_1)$ and $H(X_2)$, panel a. The generalization of mutual information to more than two variables is visualized as the intersection of all regions in the informaiton diagram, panel b. In partial information diagrams each subset $A\subseteq \{X_1,  \dots, X_N\}$ is assigned a region of size $I(A;Y)$, panels c and d.}
\label{Examples of information diagrams and partial information diagrams}
\end{figure}
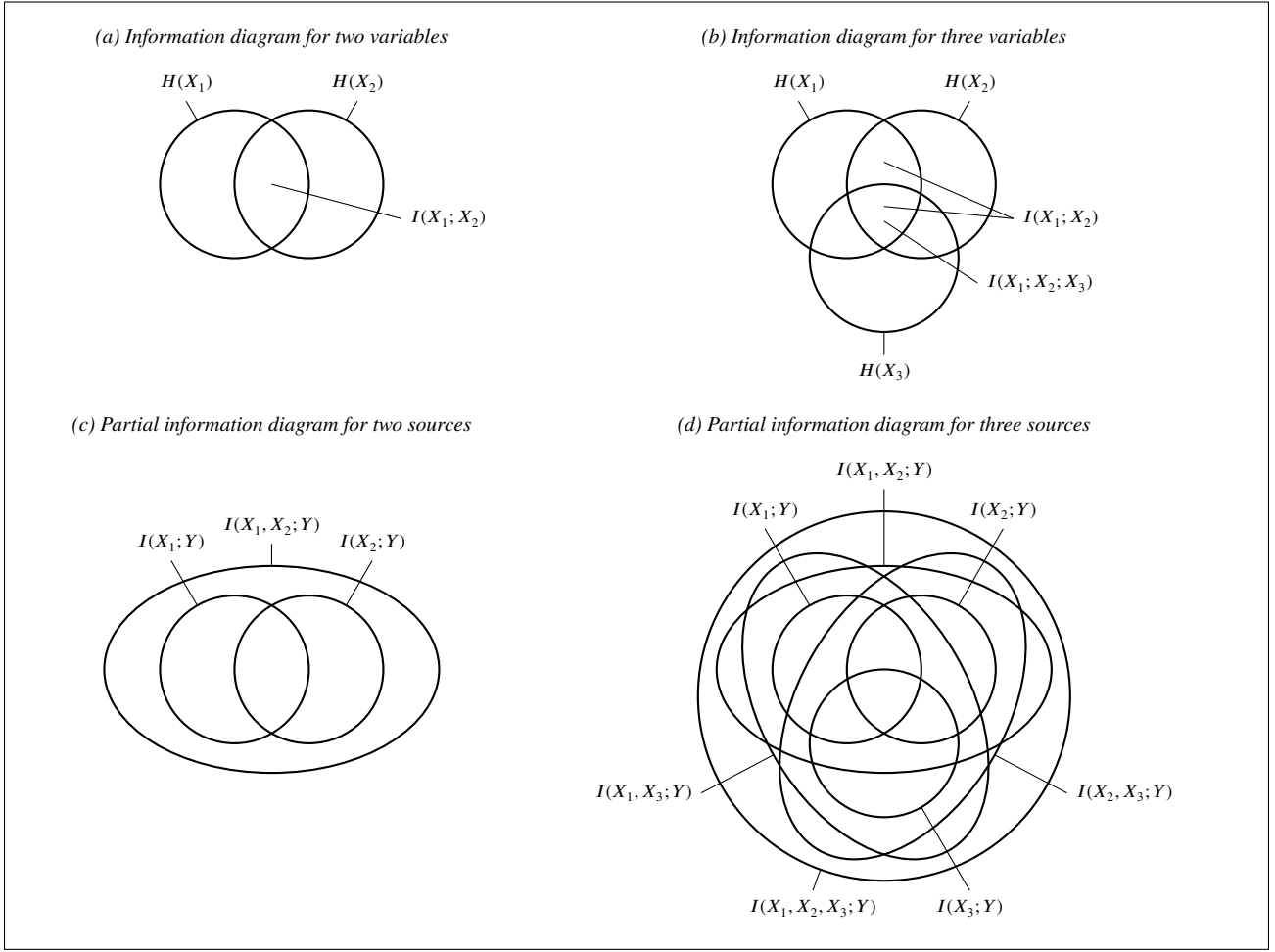

Hu and Yeung \cite{yeung_first_2002, ting_amount_1962} proposed a correspondence between information theory and set theory based on a substitution of symbols. They showed that this correspondence induces a measure (called $I$-measure) on the atoms of an information diagram, that is, a Venn diagram in which each variable is assigned a region of size corresponding to its entropy. Figures \ref{Examples of information diagrams and partial information diagrams}\emph{a} and \ref{Examples of information diagrams and partial information diagrams}\emph{b} illustrate the information diagrams for two and three variables. The problem with the Hu and Yeung approach is that it is not always clear how to interpret the $I$-measure for more than two variables, especially given that the $I$-measure can be negative.

Williams and Beer \cite{williams_nonnegative_2010} proposed a correspondence between set theory and information theory  based on the intuitive idea that a number of sources, $\{X_1,  \dots, X_N\}$, can share information about a target variable $Y$. They encoded this intuition into a number of axioms that a desirable measure of shared informaiton should satisfy. They showed that any measure of shared information that satisfies these axioms induces a measure (called PI-function) on the atoms of a partial information diagram, that is, a Venn diagram in which each subset $A\subseteq \{X_1,  \dots, X_N\}$ is assigned a region of size $I(A;Y)$. Figures \ref{Examples of information diagrams and partial information diagrams}\emph{c} and \ref{Examples of information diagrams and partial information diagrams}\emph{d} illustrate the partial information diagrams for two and three sources. Williams and Beer proposed that the PI-function has the potential to capture our intuition of synergy, redundancy and unique information. There are currently two issues with the approach proposed by Williams and Beer. The first issue \cite{kolchinsky_novel_2022} is that the proposed axioms do not identify a unique measure of shared information and, to date, no agreed-upon measure of shared information has been found. The second issue is how to relate PI-diagrams to the information diagrams of Hu and Yeung.

We proceed as follows. We introduce a novel expansion of the Shannon mutual information on  singled-out features of the target variable. We call each set of singled-out features a descriptor of the target. To address problem one, we put forward the idea that the choice of the descriptor affects the way in which the sources interact to provide the total information. We build a measure of information shared by the sources \emph{about the descriptor} and we show that this measure induces a non negative PI-function. To address problem two, we extend the descriptor-dependent measure of shared information to a measure of information shared by the sources \emph{about the target}. To address problem three, we show that the proposed measure of shared information allows linking PI-diagrams and information diagrams and allows defining a measure of multiple information that is compatible with both Hu and Yeung, and Williams and Beer set-theoretic correspondences.

\subsection{Notations and conventions}
We use uppercase letters $(X, Y, \dots)$ to indicate random variables and lowercase letters $(x, y, \dots)$ to indicate specific outcomes. $\mathbb{X}$ denotes the alphabet of $X$ and $\vert \mathbb{X} \vert$ the cardinality of $\mathbb{X}$. 
We denote with $X = \{ X_1, \dots, X_N \}$ a $N$-variate random variable and with $x = [x_1, \dots, x_N]$ it outcomes. When there is no confusion we omit the curly braces. We also denote with $\mathcal{P}(X)$ the power-set of $X$ and with $\mathcal{P}_1(X)$ the set $\mathcal{P}_1(X) = \mathcal{P}(X) \setminus \emptyset $.

We denote probability distributions with a capital letter, \emph{e.g.}, $P(X, Y)$, and values of specific realisations with lower case shorthand, \emph{e.g.}, $p(x, y)$ for $P(X{=}x, Y{=}y)$.

H(X) denotes the Shannon entropy. $I(X;Y)$ denotes the Shannon mutual information and $I(X;Y \mid Z)$ the conditional mutual information. When there is no ambiguity, we use the shorthand $I(X ; Y \mid z)$ for $I(X ; Y \mid Z{=}z)$.

\section{Expansion of mutual information on a descriptor of the target variable} \label{Sec:MI Expansion}
Let $P(X,Y)$ be a discrete probability distribution. Without implying causal relationship, we call $X$ the \emph{source} variable and $Y$ the \emph{target} variable. Consider a deterministic function $f_1: \mathbb{Y} \rightarrow \mathbb{Y}^1$. 
Because $X$ and $Y^1$ are independent given $Y$, we can rewrite Shannon's mutual information between $X$ and $Y$ \cite{shannon_mathematical_1948} as follows\footnote{We provide a setp by step derivation of Equation (\ref{MI expansion1}) in Appendix \ref{app:derivation of MI expansion1}.}
\begin{equation} \label{MI expansion1}
I (X;Y)
=
I(X;Y^1) + \sum_{y^1 \in \mathbb{Y}^1} p(y^1) \cdot I(X;Y \mid y^1)
\end{equation}
Equation (\ref{MI expansion1}) corresponds to breaking the mutual information onto different features of the target variable. The idea is that a deterministic function partitions the elements of a discrete random variable into subsets that can be interpreted as a singled out feature of $Y$ \cite{rauh_extractable_2017}. For example, let the outcomes of $Y$ be objects and let $f_1: \mathbb{Y} \rightarrow \mathbb{Y}^1$ be the function that singles out the color of an object. Each element of $\mathbb{Y}^1$ is a subset of objects of $\mathbb{Y}$ of a given color. Using Equation (\ref{MI expansion1}), $I(X;Y)$ can then be split into two parts. The first part is the average of terms of the form $I(X; Y \mid color)$, \emph{i.e.}, the information between $X$ and the elements of $Y$ of a given color. The second part is the information, $I(X; Y^1)$, conveyed by $X$ about the color variable $Y^1$.

We can further single out features from $Y^1$ through a deterministic function $f_2:\mathbb{Y}^1 \rightarrow \mathbb{Y}^2$, obtaining
\begin{equation} \label{MI expansion2}
I (X;Y)
=
I(X;Y^2) + \sum_{y^2 \in \mathbb{Y}^2} p(y^2) \cdot I(X;Y^1 \mid y^2) + \sum_{y^1 \in \mathbb{Y}^1} p(y^1) \cdot I(X;Y \mid y^1)
\end{equation}
Resuming our example with colors, let $f_2:\mathbb{Y}^1 \rightarrow \mathbb{Y}^2$ be the function that returns the temperature of a color. We can expand $I(X; Y^1)$ as the average of $I(X; Y^1 \mid warm)$ and $I(X; Y^1 \mid cool)$, plus the information, $I(X; Y^2)$, conveyed by $X$ about the temperature feature.

If we denote $Y^0 = Y$ and rewrite $I(X; Y^2)$ as $I(X;Y^2 \mid Y^3)$, where $\vert Y^3 \vert = 1$, we can write Equation (\ref{MI expansion2}) in a compact form
\begin{equation} \label{MI expansion3}
I \big( X;Y \big)
=
\sum_{\ell=1}^3
\
\sum_{y^\ell \in \mathbb{Y}^\ell}
p \big( y^\ell \big) \cdot I \big( X;Y^{\ell-1} \mid y^\ell \big)
\end{equation}
We can generalize Equation (\ref{MI expansion3}) to any deterministic Markov chain $\mathcal{Y} \triangleq Y^0 \rightarrow Y^1 \rightarrow \dotsb \rightarrow Y^L$, identified by the deterministic functions $f_\ell: \mathbb{Y}^{\ell-1} \rightarrow \mathbb{Y}^\ell$,  with $Y^0 = Y$ and $\vert \mathbb{Y}^L \vert = 1$.
Without loss of generality, we assume $\mathbb{Y}^\ell \ne \mathbb{Y}^{\ell+1}$ for all $\ell$. We obtain the expansion of $I(X;Y)$ 
\begin{equation} \label{MI expansion}
I \big( X;\mathcal{Y} \big)
\triangleq
\sum_{\ell=1}^L
\
\sum_{y^\ell \in \mathbb{Y}^\ell}
p \big( y^\ell \big) \cdot
I \big( X;Y^{\ell-1} \mid y^\ell \big) = I(X;Y).
\end{equation}
We say that $\mathcal{Y}$ is a \emph{descriptor} of $Y$ and we denote with $\Omega_Y$ the set of all possible descriptors of any length. An example of computation of Equation (\ref{MI expansion}) is shown in Figure \ref{Fig: Decompositions of UNQ}a.

We introduce two special descriptors. First, we note that the canonical expression of the Shannon mutual information \cite{shannon_mathematical_1948} corresponds to the expansion of $I(X;Y)$ obtained for the descriptor $\mathcal{S} \triangleq Y^0 \rightarrow Y^1$, with $\vert \mathbb{Y}^1 \vert = 1$. We call $\mathcal{S}$ the \emph{Shannon descriptor}. The Shannon descriptor corresponds to considering all possible features of $Y$ at once.
Second, let $Y = \{ Y_1, \dots, Y_N \}$ we introduce the \emph{canonical descriptor} $\mathcal{C}_Y = Y^0 \rightarrow \dotsb \rightarrow Y^N$ obtained using the set of functions $f^\ell(y^{\ell-1}) = [y^\ell, \dots, y^\ell_{N-\ell}] \triangleq [y^{\ell-1}_2, \dots, y^{\ell-1}_{N-\ell+1}]$. In other words, given $[y_1, \dots, y_N] \in \mathbb{Y}$, each step in the canonical chain removes one dimension, as follows, $y^1 = f^1(y) = [y_2, \dots, y_N]$, $y^2 = f^2(y^1) = [y_3, \dots, y_N]$, etc.

\section{Addressing problem one} \label{Sec: Two sources}
Let $X = \{ X_1, ..., X_N \}$ and let $A_1, \dots, A_K \in \mathcal{P}_1(X)$ be nonempty and potentially overlapping subsets of $X$, called \emph{sources} \cite{williams_nonnegative_2010}. We propose that the way the sources interact to convey the information about the target variable depends on the choice of the descriptor of $Y$. To illustrate this idea, we consider example UNQ from \cite{prokopenko_quantifying_2014}, \ref{Fig: Decompositions of UNQ}. The zero bits of shared information expected for UNQ are often explained by noticing that we can partition $\mathbb{Y}$ into two subsets. These two subsets are shown in Figure \ref{Fig: Decompositions of UNQ}\emph{a-c} and are built so that $X_1$ does not explain any of the information conveyed by $X$ \emph{within} each of the subsets. $X_2$, instead, explains all of this \emph{within} information. However, $X_2$ does not explain any of the information conveyed by $X$ \emph{between} the two subsets, while $X_1$ explains all of it. $X_1$ and $X_2$ thus convey complementary information and we expect their shared information to be null.
Indeed, for this descriptor $\mathcal{Y}$ we can write $I(X; \mathcal{Y})$ as the sum of terms that reflect the specific unique contribution of either $X_1$ or $X_2$, as shown in Figure \ref{Fig: Decompositions of UNQ}\emph{a-c}.
If we, instead, consider the Shannon descriptor, we cannot write $I(X;\mathcal{S})$ in terms of unique contributions of $X_1$ and $X_2$. We then expect the information shared by the two sources to be non-null.
This example suggests that any decomposition of the total information into the specific contribution of each possible combination of sources should be a function of the descriptor $Y$.

To build such decomposition, we follow the same approach used by Williams and Beer for constructing the $I_{min}$ measure \cite{williams_nonnegative_2010}.
We replace each term in expansion (\ref{MI expansion}) with the minimum information that any source provides about each feature of $Y$, as singled out by the descriptor $\mathcal{Y}$
\begin{equation} \label{IcapY}
I \big(A_1 {:} \dots {:} A_K; \mathcal{Y} \big)
\triangleq
\sum_{\ell=1}^L
\
\sum_{y^\ell \in \mathbb{Y}^\ell}
p \big(y^\ell\big) \cdot \min_{k=1,\dots, K} \left \{ I\big(A_k;Y^{\ell-1} \mid y^\ell\big) \right \}. 
\end{equation}
In Appendix we show that Equation (\ref{IcapY}) satisfies the Williams and Beer axioms. We also show that the Shannon descriptor maximizes Equation \ref{IcapY}.

Following the same approach used by Williams and Beers to prove that the $I_{min}$ measure induces a non-negative PI-function \cite{williams_nonnegative_2010}, it is possible to show\footnote{See also Appendix C from \cite{chicharro_synergy_2017}.} that Equation (\ref{IcapY}) induces a non-negative PI-function
\begin{equation} \label{muY}
\mu(A_1 {:} \dots {:} A_K; \mathcal{Y}) = I(A_1 {:} \dots {:} A_K; \mathcal{Y}) - \sum_{\ell=1}^{L} \ \sum_{y^\ell \in \mathbb{Y}^\ell} p(y^\ell) \cdot
\
\max_{\substack{\{ B_1, \dots, B_H \} \in \\ \{ A_1, \dots, A_K \}^-}}
\
\Big \{
\min_{h=1,\dots,H} \big \{ I(B_h; Y^{\ell-1} \mid y^\ell) \big \}
\Big \}
\end{equation}
where $\{ A_1, \dots, A_K \}^-$ denotes the subsets of $\mathcal{A}(X) := \{ \alpha \in \mathcal{P}_1(\mathcal{P}_1(X)): \forall A_i, A_j \in \alpha, A_i \centernot\subset A_j \}$, which are covered by $\{ A_1, \dots, A_K \}$ according to the redundancy partial order defined in \cite{williams_nonnegative_2010}.

While Equation (\ref{IcapY}) cannot be interpreted as a desirable measure of information shared by the sources about the target variable---because it is not univocally identified by $Y$---, it can be interpreted as a measure of the information shared by the sources about the descriptor. Accordingly, we propose to interpret Equation (\ref{muY}) as the descriptor-dependent contribution of each possible combination of the sources to the total information.
For $K=2$, Equation (\ref{muY}) provides measures of descriptor-dependent redundancy, synergy, and unique information with simple and intuitive interpretations.
The redundancy $\mu(A_1 {:} A_2; Y)$ is the minimum information conveyed by $A_1$ and $A_2$ about the features of $Y$ singled-out by $\mathcal{Y}$. The information unique to $A_1$, $\mu(A_1; Y)$, is the information about the singled-out feature of $Y$ conveyed by $A_1$ beyond the information conveyed by $X_2$. Finally te synergistic information, $\mu(A_1, A_2; Y)$ is the information about the singled-out features of $\mathcal{Y}$ that is not conveyed by either $X_1$ nor $X_2$.

\input{figures/Fig_Decompositions_of_UNQ.tex}
%

\section{Addressing problem two} \label{Sec:problem two}
We build a measure of the information shared by the sources about the target variable from Equation (\ref{IcapY}) by considering the minimum of Equation (\ref{IcapY}) over all possible descriptors of $Y$, as follows
\begin{equation} \label{Icap}
I \big(A_1{:}\dots{:} A_K; Y\big)
\triangleq \
\min_{\mathcal{Y} \in \Omega_Y}
\Big \{
I \big(A_1{:}\dots{:} A_K; \mathcal{Y})
\Big \} 
\end{equation}
The minimization removes the dependency of the shared information measure on the descriptor. We can thus interpret Equation (\ref{Icap}) as a measure of information shared about the target variable.

Equation (\ref{Icap}) satisfies several properties that have been proposed to be desirable in a measure of information shared about $Y$.
First, Equation (\ref{Icap}) returns the expected values of shared information for the canonical examples from the literature\footnote{See Appendix \ref{app:2sources:examples} for a description of these examples and their decompositions.} \cite{kolchinsky_novel_2022, prokopenko_quantifying_2014, griffith_quantifying_2015, james_multivariate_2017}.
Second, Equation (\ref{Icap}) is non-negative.
Third, Equation (\ref{Icap}), satisfies the Williams and Beer axioms\footnote{See Appendix \ref{app:2sources:identity property} for proofs of the Williams and Beer properties, the identity property, the Blackwell property, and the combined secret sharing property.}.
Fourth, Equation (\ref{Icap}) satisfies the identity property \cite{harder_bivariate_2013}.
Fifth, Equation (\ref{Icap}) satisfies the additivity property \cite{bertschinger_quantifying_2014}.
Sixth, Equation (\ref{Icap}) also satisfies the Blackwell property \cite{kolchinsky_novel_2022}.
Seventh, Equation (\ref{Icap}) satisfies the combined secret sharing property \cite{rauh_secret_2017}.
Eigth, Equation (\ref{Icap}) can be generalized to any number of sources.
Ninth, Equation (\ref{Icap}) depends only on the marginal distributions $P(A_k,Y)$ \cite{bertschinger_quantifying_2014}.
Tenth, the information quantified in Equation (\ref{Icap}) is accessible \cite{bertschinger_quantifying_2014}, since it can be extracted \cite{rauh_extractable_2017} from the realizations of $A_1, \dots, A_K$.
Finally, eleventh, as proposed in previous work \cite{kolchinsky_novel_2022}, Equation (\ref{Icap}) identifies a dual measure, $U \big(A_1 {:} \dots {:} A_K; Y \big)$, of the information jointly conveyed by the source about the target variable, as follows
\begin{equation} \label{Icup}
U \big(A_1 {:} \dots {:} A_K; Y \big)
\triangleq \
\max_{\mathcal{Y} \in \Omega_Y}
\Big \{
U \big(A_1 {:} \dots {:} A_K; \mathcal{Y})
\Big \} 
\end{equation}
where
\begin{equation} \label{IcupY}
U \big(A_1 {:} \dots {:} A_K; \mathcal{Y} \big)
\triangleq
\sum_{\ell=1}^L
\
\sum_{y^\ell \in \mathbb{Y}^\ell}
p \big(y^\ell\big) \cdot \max_{k=1,\dots, K} \left \{ U\big(A_k;Y^{\ell-1} \mid y^\ell\big) \right \}. 
\end{equation}
Equation (\ref{IcupY}) satisfies properties which are dual to those of Equation (\ref{Icap}). Equations (\ref{IcapY}) and (\ref{IcupY}) are related by the inclusion-exclusion principle through the maximum-minimums identity. However, the same is not true for Equations (\ref{Icap}) and (\ref{Icup}), except\footnote{For the case $K=2$, see the proof of the identity property, Appendix \ref{app:2sources:identity property}.} that for the case $K=2$. For $K=2$ the maximum-minimums identity guarantees that a single descriptor exists, which minimizes Equation (\ref{Icap}) for all collections of sources. For two sources, therefore the PI-function $\mu(A_1 {:} \dots {:} A_K;Y)$ induced by Equation (\ref{Icap}) is also guaranteed to be non negative, a property which has been called local positivity \cite{gilbert_shared_2013}. For $K>2$ a descriptor that minimizes Equation (\ref{Icap}) for all choices of $\{A_1, \dots, A_K \}$ does not necessarily exist. Equation (\ref{Icap}), however, is thus not guaranteed to satisfy local positivity for $K>2$. This result is in agreement with the fact that there can be no measures of shared information that satisfies the Williams and Beer axioms, the identity property and local-positivity for $K > 2$ \cite{rauh_reconsidering_2014}. In other words, for $K>2$ we might not be able to generate a PI-diagram in which all intersections can be interpreted as information shared by the sources about $Y$.

The size of $\Omega_Y$ grows according to the rate of Bell numbers. Computing Equation (\ref{Icap}) proves prohibitive for $\vert \mathbb{Y} \vert > 6$ on a normal personal computer. In appendix \ref{app:computation} we show that it is possible to considerably reduce the computation burden by restricting the domain of the minimization to the set $\Omega_L^2$ of the descriptors satisfying $\vert f_\ell^{-1}(y^\ell) \vert \le 2$ for all $y^\ell \in \mathbb{Y}^\ell$ and all $\ell = 1, \dots, L$. In other words
\begin{equation} \label{Icap}
I \big(A_1{:}\dots{:} A_K; Y\big)
= \
\min_{\mathcal{Y} \in \Omega_Y^2}
\Big \{
I \big(A_1{:}\dots{:} A_K; \mathcal{Y}).
\Big \} 
\end{equation}

\section{Addressing problem three} \label{Sec:Multivariate mutual information}

\begin{figure}[b!]
\begin{framed}
%
\begin{subfigure}[t]{0.33\textwidth}
\centering
	\subcaption[]{\centering Information diagram for three variables \newline Standard representation}
	\centering
\begin{tikzpicture}
	\def\firstcircle{ (0,0) circle (1)}
	\def\secondcircle{(1,0) circle (1)}
	\def\thirdcircle{ (0.5,-1) circle (1)}
	\begin{scope}
		\clip \firstcircle;
		\clip \secondcircle;
     	\fill[pattern=north east lines, pattern color = gray] \firstcircle;
    \end{scope}
     	
    \begin{scope}
		\clip \firstcircle \secondcircle;
     	\fill[pattern=north east lines, pattern color = gray] \thirdcircle;
     	
    		\clip \firstcircle;
    		\clip \secondcircle;
    		\clip \thirdcircle;
     	\fill[white] \thirdcircle;
     	
	\end{scope}
	
	\begin{scope}
    		\clip \firstcircle;
    		\clip \secondcircle;
    		\clip \thirdcircle;
     	\fill[pattern=north west lines, pattern color = gray] \thirdcircle;
     	\fill[pattern=north east lines, pattern color = gray] \thirdcircle;
	\end{scope}

	\node[draw, circle, minimum width = 2 cm, line width = 0.8pt] (I1) at (0,0) {};
	\node[draw, circle, minimum width = 2 cm, line width = 0.8pt] (I2) at (1,0) {};
	\node[draw, circle, minimum width = 2 cm, line width = 0.8pt] (I3) at (0.5,-1) {};
	
	\draw[] (120:1.3) -- (120:1);
	\node[anchor = south] at (120:1.3) {\scriptsize{$H(X_1)$}};

	\draw[] (1,0) +(60:1.3) -- +(60:1);
	\node[anchor = south] at ($ (1,0) +(60:1.3) $) {\scriptsize{$H(X_2)$}};

	\draw[] (0.5,-1) +(270:1.3) -- +(270:1);
	\node[anchor = north] at ($(0.5,-1) +(270:1.3)$) {\scriptsize{$H(X_3)$}};
	
	\draw[] (0.5,0) +(0,0.3) -- +(-15:1.8);
	\draw[] (0.5,0) +(0,-0.3) -- +(-15:1.8);
	\node[anchor = west] at ($(0.5,0)+(-15:1.8)$) {\scriptsize{$I(X_1;X_2)$}};
	
	\draw[] (0.5,-1) +(0,0.5) -- +(-15:1.3);
	\node[anchor = west] at ($(0.5,-1)+(-15:1.3)$) {\scriptsize{$I(X_1;X_2;X_3)$}};
		
	\end{tikzpicture}
\end{subfigure}
\hfill
\begin{subfigure}[t]{0.33\textwidth}
\centering
	\subcaption[]{\centering Information diagram for three variables \newline Blackwell's property -- part I}
	\centering
\begin{tikzpicture}
	\def\firstcircle{ (0,0) circle (1)}
	\def\secondcircle{(0.5,-1) circle (1)}
	\def\thirdcircle{ (0.4,-1) circle (0.6)}
	\begin{scope}
		\clip \firstcircle;
		\clip \secondcircle;
     	\fill[pattern=north east lines, pattern color = gray] \firstcircle;
    \end{scope}

	\begin{scope}
    		\clip \firstcircle;
    		\clip \secondcircle;
    		\clip \thirdcircle;
     	\fill[pattern=north west lines, pattern color = gray] \thirdcircle;
     	\fill[pattern=north east lines, pattern color = gray] \thirdcircle;
	\end{scope}

	\node[draw, circle, minimum width = 2 cm, line width = 0.8pt] (I1) at (0,0) {};
	\node[draw, circle, minimum width = 2 cm, line width = 0.8pt] (I2) at (0.5,-1) {};
	\node[draw, circle, minimum width = 1.2 cm, line width = 0.8pt] (I3) at (0.4,-1) {};
	
	\draw[] (120:1.3) -- (120:1);
	\node[anchor = south] at (120:1.3) {\scriptsize{$H(X_1)$}};

	\draw[] (0.4,-1) +(20:1.4) -- +(20:0.6);
	\node[anchor = west] at ($ (0.4,-1) +(20:1.4) $) {\scriptsize{$H(X_2)$}};

	\draw[] (0.5,-1) +(270:1.3) -- +(270:1);
	\node[anchor = north] at ($(0.5,-1) +(270:1.3)$) {\scriptsize{$H(X_3)$}};
	
	
	\draw[] (0.5,-1) +(-0.25,0.35) -- +(-15:1.3);
	\node[anchor = west] at ($(0.5,-1)+(-15:1.3)$) {\scriptsize{$I(X_1;X_2;X_3)$}};
	\node[anchor = west] at ($(0.5,-1.3)+(-15:1.3)$) {\scriptsize{$= I(X_1;X_2)$}};

	\end{tikzpicture}
\end{subfigure}
\hfill
\begin{subfigure}[t]{0.33\textwidth}
\centering
	\subcaption[]{\centering Information diagram for three variables \newline Blackwell's property -- part II}
	\centering
\begin{tikzpicture}
	\def\firstcircle{ (0,0) circle (1)}
	\def\secondcircle{(0,0.1) circle (0.7)}
	\def\thirdcircle{ (-0.1,0) circle (0.4)}
	\begin{scope}
		\clip \firstcircle;
		\clip \secondcircle;
     	\fill[pattern=north east lines, pattern color = gray] \firstcircle;
    \end{scope}
     	
%
%
	
	\begin{scope}
    		\clip \firstcircle;
    		\clip \secondcircle;
    		\clip \thirdcircle;
     	\fill[pattern=north west lines, pattern color = gray] \thirdcircle;
     	\fill[pattern=north east lines, pattern color = gray] \thirdcircle;
	\end{scope}

	\node[draw, circle, minimum width = 2 cm, line width = 0.8pt] (I1) at (0,0) {};
	\node[draw, circle, minimum width = 1.4 cm, line width = 0.8pt] (I2) at (0,0.1) {};
	\node[draw, circle, minimum width = 0.8 cm, line width = 0.8pt] (I3) at (-0.1,0) {};
	
	\draw[] (-0.1,0) +(120:1.3) -- +(120:0.4);
	\node[anchor = south] at ($ (-0.1,0) +(120:1.3) $) (H1) {\scriptsize{$H(X_1)$}};
	\node[anchor = south] at ($ (1.2,0) +(120:1.3) $) {\scriptsize{$= I(X_1;X_2;X_3)$}};

	\draw[] (0,0.1) +(30:1.3) -- +(30:0.7);
	\node[anchor = west] at ($ (0,0.1) +(30:1.3) $) {\scriptsize{$H(X_2)$}};

	\draw[] (0,0) +(270:1.3) -- +(270:1);
	\node[anchor = north] at ($(0,0) +(270:1.3)$) {\scriptsize{$H(X_3)$}};
	
	\draw[] (-0.1, 0) -- (-15:1.3);
	\draw[] (0.45, 0) -- (-15:1.3);
	\node[anchor = west] at (-15:1.3) {\scriptsize{$I(X_1;X_2)$}};

	\end{tikzpicture}
\end{subfigure}

\vspace*{0.5cm}

\begin{subfigure}[t]{0.33\textwidth}
\centering
	\subcaption[]{\centering Information diagram for three variables \newline Shannon's property -- part I}
	\centering
\begin{tikzpicture}
	\def\firstcircle{ (0,0) circle (1)}
	\def\secondcircle{(1,0) circle (1)}
	\begin{scope}
		\clip \firstcircle;
		\clip \secondcircle;
     	\fill[pattern=north east lines, pattern color = gray] \firstcircle;
    \end{scope}

		\node[draw, circle, minimum width = 2 cm, line width = 0.8pt] (I1) at (0,0) {};
		\node[draw, circle, minimum width = 2 cm, line width = 0.8pt] (I2) at (1,0) {};
		\node[draw, circle, minimum width = 2 cm, line width = 0.8pt] (I3) at (0.5,-2.1) {};
		
	\draw[] (120:1.3) -- (120:1);
	\node[anchor = south] at (120:1.3) {\scriptsize{$H(X_1)$}};

	\draw[] (1,0) +(60:1.3) -- +(60:1);
	\node[anchor = south] at ($ (1,0) +(60:1.3) $) {\scriptsize{$H(X_2)$}};

	\draw[] (0.5,-2.1) +(270:1.3) -- +(270:1);
	\node[anchor = north] at ($(0.5,-2.1) +(270:1.3)$) {\scriptsize{$H(X_3)$}};

	\draw[] (0.5, 0) -- ($(0.5,0)+(-15:1.8)$);
	\node[anchor = west] at ($(0.5,0)+(-15:1.8)$) {\scriptsize{$I(X_1;X_2)$}};
		
	\end{tikzpicture}
\end{subfigure}
\hfill
\begin{subfigure}[t]{0.33\textwidth}
\centering
	\subcaption[]{\centering Information diagram for three variables \newline Shannon's property - part II}
	\centering
\begin{tikzpicture}
	\node[draw, circle, minimum width = 2 cm, line width = 0.8pt] (I1) at (0,0) {};
	\node[draw, circle, minimum width = 2 cm, line width = 0.8pt] (I2) at (2.2,0) {};
	\node[draw, circle, minimum width = 2 cm, line width = 0.8pt] (I3) at (1,-1.9) {};
	
	\draw[] (120:1.3) -- (120:1);
	\node[anchor = south] at (120:1.3) {\scriptsize{$H(X_1)$}};

	\draw[] (2.2,0) +(60:1.3) -- +(60:1);
	\node[anchor = south] at ($ (2.2,0) +(60:1.3) $) {\scriptsize{$H(X_2)$}};

	\draw[] (1,-1.9) +(270:1.3) -- +(270:1);
	\node[anchor = north] at ($(1,-1.9) +(270:1.3)$) {\scriptsize{$H(X_3)$}};
		
	\end{tikzpicture}
\end{subfigure}
\hfill
\begin{subfigure}[t]{0.33\textwidth}
\centering
	\subcaption[]{\centering Information diagram for three variables \newline Shannon's property - part III}
	\centering
\begin{tikzpicture}
	\def\firstcircle{ (0,0) circle (1)}
	\def\secondcircle{(1.6,0) circle (1)}
	\def\thirdcircle{ (1,-1.57) circle (1)}
	\begin{scope}
		\clip \firstcircle;
		\clip \secondcircle;
     	\fill[pattern=north east lines, pattern color = gray] \firstcircle;
    \end{scope}
     	
    \begin{scope}
		\clip \firstcircle \secondcircle;
     	\fill[pattern=north east lines, pattern color = gray] \thirdcircle;
     	
    		\clip \firstcircle;
    		\clip \secondcircle;
    		\clip \thirdcircle;
     	\fill[white] \thirdcircle;
     	
	\end{scope}
	
	\begin{scope}
    		\clip \firstcircle;
    		\clip \secondcircle;
    		\clip \thirdcircle;
     	\fill[pattern=north west lines] \thirdcircle;
     	\fill[pattern=north east lines] \thirdcircle;
	\end{scope}

	\node[draw, circle, minimum width = 2 cm, line width = 0.8pt] (I1) at (0,0) {};
	\node[draw, circle, minimum width = 2 cm, line width = 0.8pt] (I2) at (1.6,0) {};
	\node[draw, circle, minimum width = 2 cm, line width = 0.8pt] (I3) at (1,-1.57) {};
	
	\draw[] (120:1.3) -- (120:1);
	\node[anchor = south] at (120:1.3) {\scriptsize{$H(X_1)$}};

	\draw[] (1.6,0) +(60:1.3) -- +(60:1);
	\node[anchor = south] at ($ (1.6,0) +(60:1.3) $) {\scriptsize{$H(X_2)$}};

	\draw[] (1,-1.57) +(270:1.3) -- +(270:1);
	\node[anchor = north] at ($(1,-1.57) +(270:1.3)$) {\scriptsize{$H(X_3)$}};
	
	\draw[] (0.8, 0) -- ($(0.5,0)+(-15:2.3)$);
	\node[anchor = west] at ($(0.5,0)+(-15:2.3)$) {\scriptsize{$I(X_1;X_2)$}};
		
	\end{tikzpicture}
\end{subfigure}
		
\end{framed}
\caption{Information diagrams for three variables. a) Standard representation. Each variable is assigned a region of size corresponding to its entropy. The multiple mutual information is visualized as the intersection of the three regions. b) If $X_2 = f(X_3)$ we expect $I(X_1;X_2;X_3) = I(X_1;X_2)$. c) If $X_2 = f(X_3)$ and $X_1 = f'(X_2)$ we expect $I(X_1;X_2;X_3) = H(X_1)$. d) If $p(x_1,x_3) = p(x_1) \cdots p(x_3)$ and $p(x_2, x_3) = p(x_2) \cdot p(x_3)$ we expect $I(X_1;X_2;X_3) = 0$. e) If $p(x_1,x_2,x_3) = p(x_1) \cdot p(x_2) \cdot p(x_3)$ we expect $I(x_1;X_2;X_3)=0$. f) We expect that it is possible to have $I(X_1;X_2;X_3) = 0$ even when $I(X_1;X_2), I(X_2;X_3), I(X_1;X_3) > 0$.}
\label{Fig:PID_2S}
\end{figure}
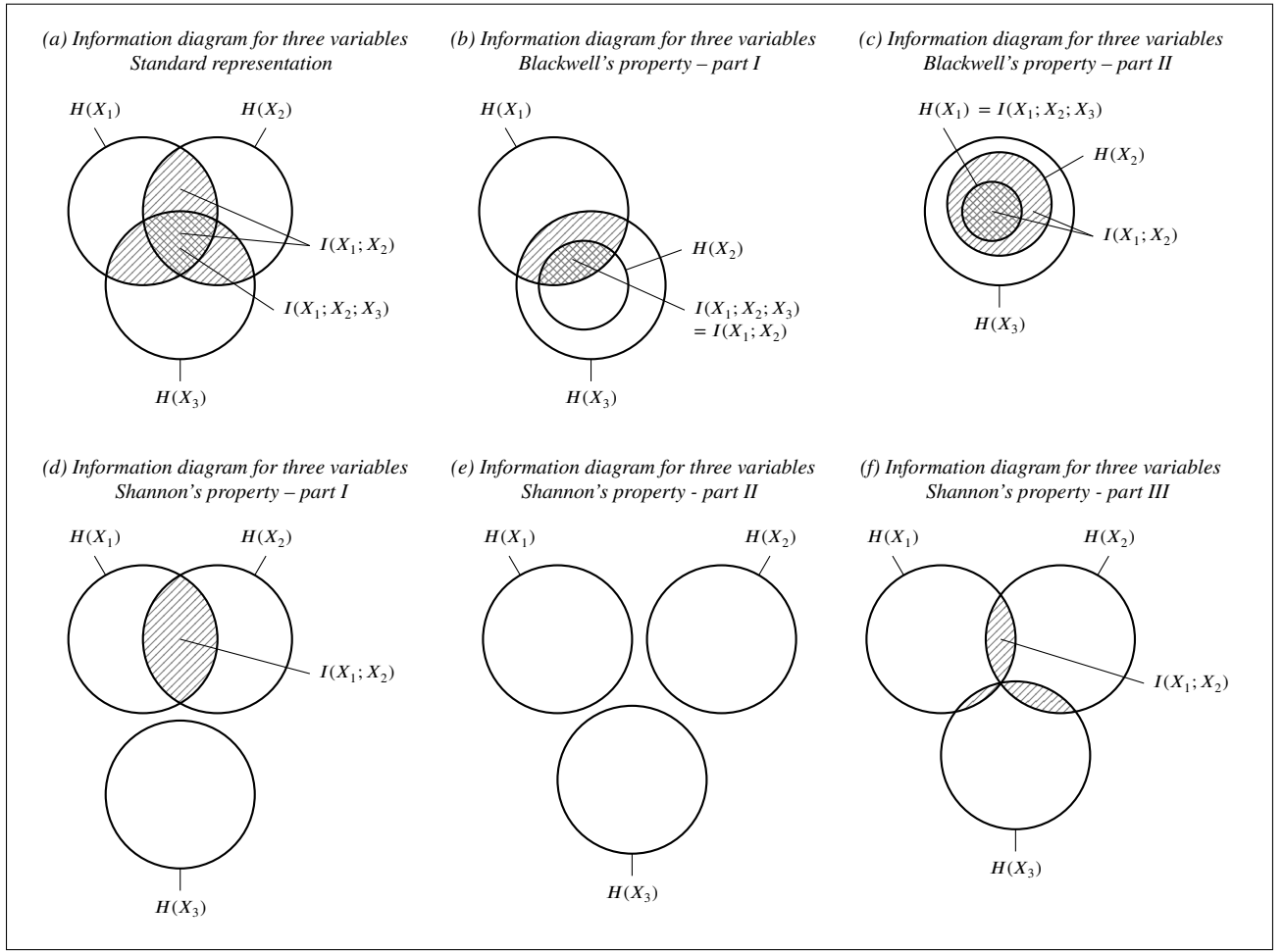

\begin{figure}[t!]
\begin{framed}
\begin{subfigure}[t]{0.4\textwidth}
\centering
	\subcaption[short for lof]{Partial information diagram for two sources}
	\centering
\begin{tikzpicture}
		
		\node[draw, circle, minimum width = 2 cm, line width = 0.8pt] (I1) at (0,0) {};
		\node[draw, circle, minimum width = 2 cm, line width = 0.8pt] (I2) at (1,0) {};
		\node[draw, ellipse, minimum width = 4.5 cm, minimum height = 2.8 cm, line width = 0.8pt] (I12) at (0.5,0) {};
		
		
		\draw[] (0.5, 1.8) -- (I12.north);
		\node[anchor = south] at (0.5,1.8) {\scriptsize{$I(X_1,X_2;Y)$}};
		
		
%
		
		
		\draw[] (0,0) +(120:1.7) -- +(120:1);
		\node[anchor = south] at ($ (0,0) +(120:1.7) $) {\scriptsize{$I(X_1;Y)$}};
		
		\draw[] (1,0) +(60:1.7) -- +(60:1);
		\node[anchor = south] at ($ (1,0) +(60:1.7) $) {\scriptsize{$I(X_2;Y)$}};
		
		\node[anchor = center] () at (0.5,0) {\scriptsize{$1\text{:}2$}};
		\node[anchor = center] () at (-0.4,0) {\scriptsize{$1$}};
		\node[anchor = center] () at (1.4,0) {\scriptsize{$2$}};
		\node[anchor = center] () at (0.5,1.15) {\scriptsize{$12$}};
		
	\end{tikzpicture}
\end{subfigure}
\begin{subfigure}[t]{0.6\textwidth}
\centering
	\subcaption[short for lof]{Partial information diagram for two sources and $Y = \{ X_1, X_2 \}$}
\begin{tikzpicture}
		
		\node[draw, circle, minimum width = 2 cm, line width = 0.8pt] (I1) at (0,0) {};
		\node[draw, circle, minimum width = 2 cm, line width = 0.8pt] (I2) at (1,0) {};
		
		\draw[line width = 1.2pt,domain=62:298,color=gray, dashed] plot ({1.09*cos(\x)}, {1.09*sin(\x)});
		\draw[line width = 1.2pt,domain=-118:118,color=gray, dashed] plot ({1.09*cos(\x)+1}, {1.09*sin(\x)});
		
		\draw[line width = 1.2pt, color = gray, dashed] (0.5, 1.8) -- (0.5, 0.97);
		\node[anchor = south] (I12) at (0.5, 1.8) {\scriptsize{$I(X_1,X_2;Y)$}};
		\node[right = -0.12cm of I12, anchor = west] () {\scriptsize{$= H(X_1,X_2)$}};

		\draw[] (0,0) +(120:1.4) -- +(120:1);
		\node[anchor = south] (I1) at ($ (0,0) +(120:1.4) $) {\scriptsize{$I(X_1;Y)$}};
		\node[left = -0.14cm of I1, anchor = east] () {\scriptsize{$H(X_1)=$}};
		
		\draw[] (1,0) +(60:1.4) -- +(60:1);
		\node[anchor = south] (I2) at ($ (1,0) +(60:1.4) $) {\scriptsize{$I(X_2;Y)$}};
		\node[right = -0.14cm of I2, anchor = west] () {\scriptsize{$= H(X_2)$}};
		
		
		

	\draw[] (0.5,0) +(0,-0.3) -- +(-15:1.8);
	\node[anchor = west] at ($(0.5,0)+(-15:1.8)$) {\scriptsize{$I(X_1;X_2)$}};
		
		\node[anchor = center] () at (0.5,0) {\scriptsize{$1\text{:}2$}};
		\node[anchor = center] () at (-0.4,0) {\scriptsize{$1$}};
		\node[anchor = center] () at (1.4,0) {\scriptsize{$2$}};

	\end{tikzpicture}
	\end{subfigure}
		
\end{framed}
\caption{Partial information diagrams for two sources for the measure of shared information defined in Equation (\ref{Icap}). We use the shorthand notation $i{:}j$ for $\mu(X_i{:}X_j;Y)$. a) Standard representation. b) If $Y=X$ the partial information diagram for two sources reduces to the information diagram for two variables.}
\label{Fig:PID_2S_copy}
\end{figure}
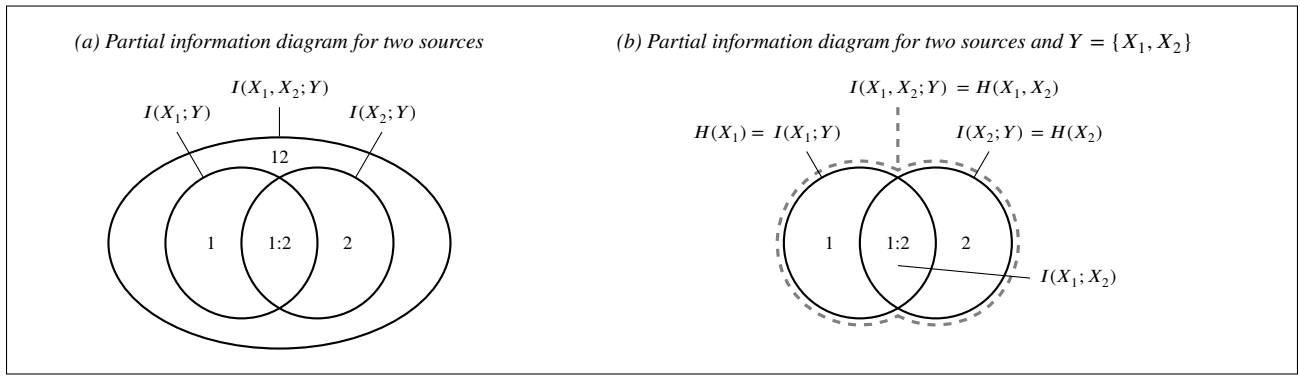

The correspondence between Shannon's information measures and information diagrams for two variables has led to hypothesize the existence of a generalization of the Shannon mutual information to more than two variables \cite{yeung_first_2002}, Figure \ref{Fig:PID_2S}\emph{a}. Currently, no agreed-upon measure of multiple information has been identified \cite{timme_synergy_2014}. Based on the information-diagram correspondence, we propose that a measure, $I(X_1; \dots; X_N)$, of multiple information should satisfy the following properties:
\begin{enumerate}
\item
\emph{Non-negativity}: $I(X_1; \dots; X_N) \ge 0$.
\item
\emph{Symmetry}: $I(X_1; \dots; X_N)$ is invariant to permutations of $X_1, \dots, X_N$.
\item
\emph{Monotonicity}: $I(X_1; \dots; X_{N-1}) \ge I(X_1; \dots ; X_N)$.
\item
\emph{Self-information}: $I(X_1; \dots; X_N)$ reduces to the Shannon mutual information and the Shannon entropy for $N=2$ and $N=1$, respectively.
\item
\emph{Blackwell property}: If $X_{N-1} = f(X_N)$ then $I(X_1; \dots; X_N) = I(X_1; \dots; X_{N-1})$, Figure \ref{Fig:PID_2S}\emph{b}. This also implies $I(X_1; \dots; X_N) = H(X_1)$ if $X_1, \dots, X_N$ form a Markov chain $X_N \rightarrow X_{N-1} \rightarrow \dots \rightarrow X_1$, Figure \ref{Fig:PID_2S}\emph{c}.
\item
\emph{Shannon property}: If $\bar{n}\in \{ 1, \dots, N \}$ exists such that $p(x_{\bar{n}}, x_n) = p(x_{\bar{n}}) \cdot p(x_n)$, for all $n \ne \bar{n}$ and all $x \in X$, then $I(X_1; \dots; X_N) = 0$, Figure \ref{Fig:PID_2S}\emph{d}. This also implies $I(X_1; \dots; X_N) = 0$ if $p(x_1, \dots, x_N) = p(x_1) \cdot \dots \cdot p(x_N)$ for all $x \in X$, Figure \ref{Fig:PID_2S}\emph{e}. However, $I(X_1; \dots; X_N)=0$ does \emph{not} imply $p(x_{n_1}, x_{n_2}) = p(x_{n_1}) \cdot p(x_{n_2})$ for any $n_1, n_2 = 1, \dots, N$, Figure \ref{Fig:PID_2S}\emph{f}.
\end{enumerate}
Properties 2-4 are reminiscent of the Williams and Beer axioms for shared information \cite{williams_nonnegative_2010}. This correspondence suggests that it might be possible to derive a measure of multiple information from a measure of shared information.
To this aim we note that, for any measure of shared information that satisfies the identity property, the PI-diagram for two sources, Figure \ref{Fig:PID_2S_copy}\emph{a}, reduces to to the information diagram for two variables when $Y = X$, Figure \ref{Fig:PID_2S_copy}\emph{a}. Based on this correspondence, we propose to define multiple information $I(X_1; \dots, ; X_N)$, as follows
\begin{equation} \label{multipleMI}
I(X_1; \dots, ; X_N) \triangleq I(X_1 {:} \dots {:} X_N; X_1, \dots, X_N)
\end{equation}
Local-positivity, symmetry, monotonicity, self-information and the Blackwell property follow from the properties of Equation (\ref{Icap}). In Appendix \ref{app:proof property 4}, we also show that Equation (\ref{multipleMI}) satisfies the Shannon property.

\section{Discussion}
The measures of descriptor-dependent and descriptor-independent shared information proposed in this work are a direct extention of the approach proposed by Williams and Beer. Instead of minimizing the terms of a point-wise decompoeision, we minimize the terms of a novel decomposition of mutual information, Equation (\ref{MI expansion}). Unlike point-wise decompositions \cite{deweese_how_1999}, Equation (\ref{MI expansion}) does not attempt to decompose mutual information into singled-out \emph{outcomes} of the target variable. Instead, Equation (\ref{MI expansion}) decomposes mutual information onto singled-out \emph{features} of the target variable. Mutual information is intrisically a non point-wise measure, as epitomized by the fact that mutual information is null whenever the alphabet of any of the two arguments has cardinality one. While the terms of point-wise decompositions cannot be interpreted in terms of Shannon's information quantities, the terms of Equation (\ref{MI expansion}) are themselves mutual informations.

We proposed that, for more than two sources, problem one should be disentangled from problem two. The problem of quantifying the information conveyed specifically by a collection of sources depends on the choice of the descriptor of the target variable. Instead, shared information is descriptor-independent. Our approach allows reconciling our intuition that the information conveyed specifically by a collection of sources should be non-negative with the results from \cite{rauh_reconsidering_2014} that a non-negative measure of shared information is not compatible with the partial information decomposition. An important open research question will be to identify the sufficient conditions that $P(X,Y)$, $N>2$,  must satisfy to ensure that a single descriptor exists which minimizes equation (\ref{Icap}) for all choiches of the collection of sources.

To our knowledge, our measure of shared information is the only proposed measure satisfying the ten highlighted properties. It is easy to show\footnote{See Theorem 6 in Appendix \ref{app:proof property 4} for an example.} that our measure does not satisfy left monotonicity \cite{gilbert_shared_2013}. We propose that no measure of shared information exists, which satisfies left monotonicity and is compatible with the accepted values of shared information for the canonical examples. This is because distribution\footnote{See Appendix \ref{app:2sources:examples} for a description of examples \textsc{And} and \textsc{Unq} and their decompositions.} \textsc{And} can be obtained from distribution \textsc{Unq} \cite{prokopenko_quantifying_2014} through a transformation of the realizations of the target. However the information shared by $X_1$ and $X_2$ in \textsc{And} is expected to be higher than that in \textsc{Unq} \cite{prokopenko_quantifying_2014}. This also implies that no measure of shared information exists which is compatible with the expected breakdown for the canonical examples and which satisfies strong symmetry and the left chain rule \cite{gilbert_shared_2013}.

We proposed a new measure of multiple information. We note that among the measures of multiple information proposed in the literature \cite{timme_synergy_2014} ours and McGill's interaction information \cite{mcgill_multivariate_1954, yeung_first_2002} are compatible with the set-theoretic intuition of multiple information that we derive from the information diagrams. However, unlike interaction information \cite{williams_nonnegative_2010, yeung_first_2002}, our measure of multiple information is guaranteed to be non-negative.


\subsection{Acknowledgements}
I would like to thank Andrei Romashchenko for providing a counterexample, which was fundamental to the development of this theory, Daniel Chicharro and Artemy Kolchinsky for answering many of my questions on the partial information decomposition theory, and Niklas Ludtke for his feedback to the manuscript.

\bibliography{Library}
\bibliographystyle{unsrt}

\newpage
\appendix
\section{Supplemental information and supporting proofs} \label{App:Supplementary material}

\subsection{Step-by-step derivation of Equation (\ref{MI expansion1})} \label{app:derivation of MI expansion1}

\begin{theorem}
Consider $P(X,Y)$ discrete and $f: \mathbb{Y} \rightarrow \mathbb{Y}^1$ deterministic. We have $I(X;Y) = I(X ; Y^1) + I(X;Y \mid Y^1)$.
\end{theorem}

\begin{proof}
Remember that any deterministic function $f:\mathbb{Y} \rightarrow \mathbb{Y}^1$ partitions the elements of $\mathbb{Y}$ into subsets that correspond to the element of the $\mathbb{Y}^1$. That means that for $y^1 \in \mathbb{Y}^1$ we have $y^1 \subseteq \mathbb{Y}$. From the definition of the Shannon mutual information, we have
\begin{align*}
I(X;Y)
& =
\sum_{\subalign{x &\smallin \mathbb{X} \\ y &\smallin \mathbb{Y}}} p(x, y) \cdot \log \frac{p(x, y)}{p(x) \cdot p(y)} =
%
H(X) \ + \ \smashoperator{\sum_{\subalign{x &\smallin \mathbb{X} \\ y &\smallin \mathbb{Y}}}} p(x, y) \cdot \log \frac{p(x, y)}{p(y)} = \\
&
\overset{(a)}{=} H(X) \ + \ \smashoperator{\sum_{\subalign{x &\smallin \mathbb{X} \\ y^1 &\smallin \mathbb{Y}^1 \\ y &\smallin y^1}}} \ \frac{p(x, y)}{r(y, Y^1)} \cdot \log \frac{p(x, y)}{p(y)}
\overset{(b)}{=} H(X) \ + \ \smashoperator{\sum_{\subalign{x &\smallin \mathbb{X} \\ y &\smallin \mathbb{Y} \\ y^1 &\smallin \mathbb{Y}^1}}}
p(x, y, y^1) \cdot \log \frac{P(x, y \mid y^1)}{P(y \mid y^1)} \\
&
\overset{(c)}{=}
H(X) - H(X \mid Y^1) + I(X;Y \mid Y^1)
=
I(X ; Y^1) + I(X;Y \mid Y^1)
\end{align*}
where in (a) $r(y, Y^1)$ denotes the number of elements of $Y^1$ that include $y$. For (b) we used the relationship $p(x, y) / p(y) = p(x, y \mid y^1) / p(y \mid y^1)$. In (c) we added and subtracted $H(X \mid Y^1)$.
\end{proof}

\subsection{Properties of Equation (\ref{IcapY})} \label{app:2sources:properties IcapY}

\begin{theorem}[Williams and Beer axioms]
Equation (\ref{IcapY}) satisfies the following properties:
\begin{enumerate}
\item
\emph{Symmetry}: $I(A_1 {:} \dots {:} A_K; \mathcal{Y})$ does not depend on the order of $A_1, \dots, A_K$
\item
\emph{Self redundancy}: $I(A {:} A; \mathcal{Y}) = I(A;\mathcal{Y})$
\item
\emph{Monotonicity}: $I(A_1 {:} \dots {:} {A_K}; \mathcal{Y}) \le I(A_1 {:} \dots {:} {A_{K-1}}; \mathcal{Y})$
\end{enumerate}
\end{theorem}
\begin{proof}
Property one follows from the the fact that the minimum operator $\min \{ a_1, \dots, a_K \}$ does not depend on the order of the elements of $\{ a_1, \dots, a_K \}$. Property two follows from the fact that equation $I(A {:} A; \mathcal{Y})$ reduces to decomposition (\ref{MI expansion}) for any choice of descriptor of $\mathcal{Y}$. We define
$$
\mathcal{Y}' \triangleq \argmin_{\mathcal{Y} \in \Omega_Y}
\Big \{
I \big(A_1{:}\dots{:} A_K; \mathcal{Y})
\Big \}
\quad
\text{and}
\quad
\mathcal{Y}'' \triangleq \argmin_{\mathcal{Y} \in \Omega_Y}
\Big \{
I \big(A_1{:}\dots{:} A_{K-1}; \mathcal{Y})
\Big \}.
$$
Because $\min \{a_1, \dots, a_K \} \le \min \{a_1, \dots, a_{K-1} \}$ for any set $\{a_1, \dots, a_K\}$, we have
$$
I \big(A_1{:}\dots{:} A_K; Y) = I \big(A_1{:}\dots{:} A_K; \mathcal{Y}') \le I \big(A_1{:}\dots{:} A_K; \mathcal{Y}'') \le I \big(A_1{:}\dots{:} A_{K-1}; \mathcal{Y}'') = I \big(A_1{:}\dots{:} A_{K-1}; Y)
$$
This proves property three.
\end{proof}

We can now prove the following theorem on the upper bound of Eqution (\ref{IcapY}).
\begin{theorem}
For any $\mathcal{Y} \in \Omega_Y$ the relationship $I(A_1 {:}\dots{:} A_K; \mathcal{Y}) \le I(A_1 {:} \cdot A_K; \mathcal{S})$ holds
\end{theorem}

\begin{proof}
Suppose that $I \big(A_1{:}\dots{:} A_{K}; \mathcal{S}) < I \big(A_1{:}\dots{:} A_{K}; \mathcal{Y})$. 
By construction, $I \big(A_1{:}\dots{:} A_{K}; \mathcal{S}) = I(A_k; \mathcal{S}) = I(A_k; Y)$ for some $k \in \{ 1, \dots, K \}$. Because $I(A_k; Y) = I(A_k; \mathcal{Y})$, we have $I(A_k; \mathcal{Y}) < I \big(A_1{:}\dots{:} A_{K}; \mathcal{Y})$ which violates the monotonicity rule.
\end{proof}

\subsection{Refining a descriptor reduces the information shared by the sources} \label{app:computation}

Let $\mathcal{Y} = Y^0 \xrightarrow{f_1} \dotsb \xrightarrow{f_{\ell-1}} Y^{\ell-1} \xrightarrow{f_\ell}  Y^\ell \xrightarrow{f_{\ell+1}} \dotsb \xrightarrow{f_L} Y^L \in \Omega_Y$. We build a new descriptor $\mathcal{Y}^* \in \Omega_Y$ from $\mathcal{Y}$, as follows: $\mathcal{Y}^* = Y^0 \xrightarrow{f_1} \dotsb \xrightarrow{f_{\ell-1}} Y^{\ell-1} \xrightarrow{f_a}  Y^* \xrightarrow{f_b} Y^\ell \xrightarrow{f_{\ell+1}} \dotsb \xrightarrow{f_L} Y^L$ with $f_a \circ f_b = f_\ell$. We say that $\mathcal{Y}^*$ is a \emph{refinement} of $\mathcal{Y}$. An example of refinement is shown in Figure \ref{Fig: Example of refinement}.

\begin{theorem} [Refining a descriptor reduces the information shared by the sources] \label{refinement theorem}
Let $\mathcal{Y}^*$ be a \emph{refinement} of $\mathcal{Y} \in \Omega_Y$, then
$$
I(A_1 {:} \dots {:} A_K; \mathcal{Y}^*) \le I(A_1 {:} \dots {:} A_K; \mathcal{Y}).
$$
\end{theorem}

\begin{proof}
Let $\bar{k} \in \argmin_{k=1,\dots,K} \left \{ I\big(A_k;Y^{\ell-1} \mid y^\ell\big) \right \}.$ Applying Equation (\ref{MI expansion1}) we have have
\begin{alignat*}{3}
& I(A_{\bar{k}}; Y^{\ell-1} \vert y^\ell) 
&& = I(A_{\bar{k}}; Y^* \vert y^\ell) &&+ \sum_{y^* \smallin f_b^{-1} \left( y^\ell \right)} I(A_{\bar{k}}; Y^{\ell-1} \vert y^*, y^\ell)  \\
& && \ge \min_{k=1,\dots,K} \left \{ I(A_k; Y^* \vert y^\ell) \right \} &&+ \sum_{y^* \smallin f_b^{-1} \left(y^\ell \right)} \min_{k=1,\dots,K} \left \{ I(A_k; Y^{\ell-1} \vert y^*, y^\ell) \right \}
\end{alignat*}
By construction we have
\begin{align*}
& I(A_1 {:} \dots {:} A_K; \mathcal{Y}) - I(A_1 {:} \dots {:} A_K; \mathcal{Y}^*) = \\
& = I(A_{\bar{k}}; Y^{\ell-1} \vert y^\ell) -
\left( \min_{k=1,\dots,K} \left \{ I(A_k; Y^* \vert y^\ell) \right \}
+  \sum_{y^* \smallin f_b^{-1}(y^\ell)} \min_{k=1,\dots,K} \left \{ I(A_k; Y^{\ell-1} \vert y^*, y^\ell) \right \} \right) \ge 0
\end{align*}
The theorem is proved.
\end{proof}
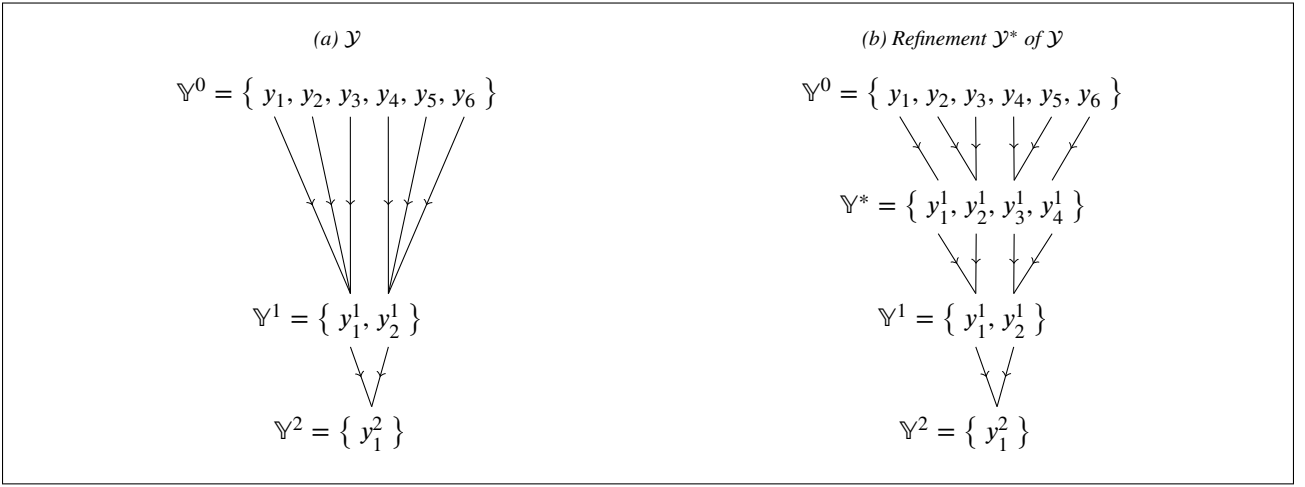
\begin{figure}[t!]
\begin{framed}
	
	\begin{subfigure}[t]{0.5\textwidth}
	\centering
	\subcaption[short for lof]{$\mathcal{Y}$}
		\begin{tikzpicture}[remember picture]
		noname/.style={%
    	rectangle, rounded corners=9pt,
    	fill=black!20}
		
		\node [anchor=west] (Y0) at (0, 3) {$\mathbb{Y}^0 = \big\{$
			\subnode{ay-0-1}{$y_1$},
			\subnode{ay-0-2}{$y_2$},
			\subnode{ay-0-3}{$y_3$},
			\subnode{ay-0-4}{$y_4$},
			\subnode{ay-0-5}{$y_5$},
			\subnode{ay-0-6}{$y_6$}
			\subnode{brkt0}{$ \big\}$}};
			
		\node [anchor=west] (Y1) at (1, 0) {$\mathbb{Y}^1 = \big\{$
			\subnode{ay-1-1}{$y^1_1$},
			\subnode{ay-1-2}{$y^1_2$}
			\subnode{brkt1}{$ \big\}$}};
			
		\node [anchor=west] (Y2) at (1.28, -1.5) {$\mathbb{Y}^2 = \big\{$
			\subnode{ay-2-1}{$y^2_1$}
			\subnode{brkt2}{$ \big\}$}};
		
		\begin{scope}[decoration={markings, mark=at position 0.5 with {\arrow{>}}}] 
			\draw[postaction={decorate}] (ay-0-1.south) -- (ay-1-1.north);
			\draw[postaction={decorate}] (ay-0-2.south) -- (ay-1-1.north);
			\draw[postaction={decorate}] (ay-0-3.south) -- (ay-1-1.north);
			\draw[postaction={decorate}] (ay-0-4.south) -- (ay-1-2.north);
			\draw[postaction={decorate}] (ay-0-5.south) -- (ay-1-2.north);

			\draw[postaction={decorate}] (ay-0-6.south) -- (ay-1-2.north);
			\draw[postaction={decorate}] (ay-1-1.south) -- (ay-2-1.north);
			\draw[postaction={decorate}] (ay-1-2.south) -- (ay-2-1.north);
		\end{scope}
		\end{tikzpicture}
	\end{subfigure}
\hfill
\begin{subfigure}[t]{0.5\textwidth}
	\centering
	\subcaption[short for lof]{Refinement $\mathcal{Y}^*$ of $\mathcal{Y}$}
		\begin{tikzpicture}[remember picture]
		noname/.style={%
    	rectangle, rounded corners=9pt,
    	fill=black!20}
		
		\node [anchor=west] (Y0) at (0, 3) {$\mathbb{Y}^0 = \big\{$
			\subnode{by-0-1}{$y_1$},
			\subnode{by-0-2}{$y_2$},
			\subnode{by-0-3}{$y_3$},
			\subnode{by-0-4}{$y_4$},
			\subnode{by-0-5}{$y_5$},
			\subnode{by-0-6}{$y_6$}
			\subnode{brkt0}{$ \big\}$}};
			
		\node [anchor=west] (Y1) at (0.5, 1.5) {$\mathbb{Y}^*= \big\{$
			\subnode{by-1-1}{$y^1_1$},
			\subnode{by-1-2}{$y^1_2$},
			\subnode{by-1-3}{$y^1_3$},
			\subnode{by-1-4}{$y^1_4$}
			\subnode{brkt1}{$ \big\}$}};
			
		\node [anchor=west] (Y1) at (1, 0) {$\mathbb{Y}^1 = \big\{$
			\subnode{by-2-1}{$y^1_1$},
			\subnode{by-2-2}{$y^1_2$}
			\subnode{brkt1}{$ \big\}$}};
			
		\node [anchor=west] (Y2) at (1.28, -1.5) {$\mathbb{Y}^2 = \big\{$
			\subnode{by-3-1}{$y^2_1$}
			\subnode{brkt2}{$ \big\}$}};
		
		\begin{scope}[decoration={markings, mark=at position 0.5 with {\arrow{>}}}] 
			\draw[postaction={decorate}] (by-0-1.south) -- (by-1-1.north);
			\draw[postaction={decorate}] (by-0-2.south) -- (by-1-2.north);
			\draw[postaction={decorate}] (by-0-3.south) -- (by-1-2.north);
			\draw[postaction={decorate}] (by-0-4.south) -- (by-1-3.north);
			\draw[postaction={decorate}] (by-0-5.south) -- (by-1-3.north);
			\draw[postaction={decorate}] (by-0-6.south) -- (by-1-4.north);
			
			\draw[postaction={decorate}] (by-1-1.south) -- (by-2-1.north);
			\draw[postaction={decorate}] (by-1-2.south) -- (by-2-1.north);
			\draw[postaction={decorate}] (by-1-3.south) -- (by-2-2.north);
			\draw[postaction={decorate}] (by-1-4.south) -- (by-2-2.north);
			
			\draw[postaction={decorate}] (by-2-1.south) -- (by-3-1.north);
			\draw[postaction={decorate}] (by-2-2.south) -- (by-3-1.north);

		\end{scope}
		\end{tikzpicture}
	\end{subfigure}

\end{framed}
\caption{Example of refinement of a descriptor. $\mathcal{Y}^*$ belongs to $\Omega_Y^2$, while $\mathcal{Y}$ only belongs to $\Omega_Y$.}
\label{Fig: Example of refinement}
\end{figure}

Theorem \ref{refinement theorem} has the following implications.

\begin{corollary}
For any $\mathcal{Y} \in \Omega_Y$ we have
$$
I(A_1 {:} \dots {:} A_K; \mathcal{Y}) \le I(A_1 {:} \dots {:} A_K; \mathcal{S})
$$
\end{corollary}

\begin{proof}
The proof follows from the fact that any $\mathcal{Y} \in \Omega_Y$ with $\mathcal{Y} \ne \mathcal{S}$ can be viewed as a refinement of $\mathcal{S}$.
\end{proof}

\begin{corollary}
The following equality holds
$$
\min_{\mathcal{Y} \in \Omega_Y}
\Big \{
I \big(A_1{:}\dots{:} A_K; \mathcal{Y}).
\Big \}
=
\min_{\mathcal{Y} \in \Omega_Y^2}
\Big \{
I \big(A_1{:}\dots{:} A_K; \mathcal{Y}).
\Big \} 
$$
\end{corollary}

\begin{proof}
Let $\mathcal{Y} \in \argmin_{\mathcal{Y} \in \Omega_Y} \Big \{ I \big(A_1{:}\dots{:} A_K; \mathcal{Y}) \Big \}$ with $\mathcal{Y} \notin \Omega_Y^2$. We can always build $\mathcal{Y}^* \in \Omega_Y^2$ by refining $\mathcal{Y}$. Theorem \ref{refinement theorem} guarantees that $\mathcal{Y}^* \in \argmin_{\mathcal{Y} \in \Omega_Y} \Big \{ I \big(A_1{:}\dots{:} A_K; \mathcal{Y}) \Big \}$. 
\end{proof}

\subsection{Two-sources examples} \label{app:2sources:examples}
In Section \ref{Sec:problem two} we showed that for the case $X = \{X_1, X_2\}$ the terms, $\mu(X_1 {:} X_2; Y)$, $\mu(X_1; Y)$, $\mu(X_2; Y)$, $\mu(X_1, X_2; Y)$, of the PI-function induced by Equation (\ref{Icap}) are guaranteed to be non-negative. Williams and Beer proposed that for a desirable measure of shared information we should be able to interpret these terms as the redundant information, the information unique to $X_1$, the information unique to $X_2$ and the synergistic information about $Y$, respectively.
In Table \ref{Tab:two sources examples} we show that Equation (\ref{Icap}) satisfies the interpretation proposed by Williams and Beer by summarizing the values obtained with our decomposition for the canonical examples from the literature \cite{kolchinsky_novel_2022, prokopenko_quantifying_2014, griffith_quantifying_2015, james_multivariate_2017}.

\begin{table}[h!]
\centering
\begin{tabular}[t]{ | l c  c  c  c  c |}
	\hline
	\rule{0pt}{11pt} & $I(X_1,X_2;Y)$ & $\mu(X_1{:} X_2;Y)$ & $\mu(X_1;Y)$ & $\mu(X_2;Y)$ & $\mu(X_1,X_2;Y)$\\
	& \footnotesize{total} & \footnotesize{redundant} & \footnotesize{unique to $X_1$} & \footnotesize{unique to $X_2$} & \footnotesize{synergistic}\\
	\hline
	\textsc{Rdn} & 1 &  1 & 0 & 0 & 0 \\
	\hline
	\textsc{Imperfect Rdn} & 1  & 0.93 & 0 & 0.07 & 0 \\
	\hline
	\textsc{Unq1} & 1 &  0 & 1 & 0 & 0 \\
	\hline
	\textsc{Unq2} & 1 &  0 & 0 & 1 & 0 \\
	\hline
	\textsc{Unq} & 2 &  0 & 1 & 1 & 0 \\
	\hline
	\textsc{Syn} & 1 &  0.5 & 0 & 0 & 0.5 \\
	\hline
	\textsc{Corner} & 0.92 &  0.25 & 0 & 0 & 0.67 \\
	\hline
	\textsc{Xor} & 1 &  0 & 0 & 0 & 1 \\
	\hline
	\textsc{And} & 0.81 &  0.31 & 0 & 0 & 0.5 \\
	\hline
	\textsc{Sum} & 1.5 & 0.5 & 0 & 0 & 1 \\
	\hline
	\textsc{Dyadic} & 2 & 0  & 1 & 1 & 0 \\
	\hline
	\textsc{Triadic} & 2 & 1 & 0 & 0 & 1 \\
	\hline
	\textsc{RdnXor} & 2 & 1 & 0 & 0 & 1 \\
	\hline
	\textsc{RdnUnqXor} & 4 &  1 & 1 & 1 & 1 \\
	\hline
	\hline
	$Y = X_1$ & \small{$I(X_1, X_2; X_1)$} & \small{$I(X_1; X_2)$} & \small{$I(X_1; Y) - I(X_1;X_2)$} & 0 & 0 \\
	\hline
	$Y = \{ X_1, X_2 \}$ & \small{$I(X_1, X_2; \{ X_1, X_2 \})$} & \small{$I(X_1; X_2)$} & \small{$I(X_1; Y) - I(X_1;X_2)$} & \small{$I(X_2; Y) - I(X_1;X_2)$} & 0 \\
	\hline
\end{tabular}
\caption{Values of total information, $I(X_1, X_2;Y)$, redundant information, $\mu(X_1 {:} X_2;Y)$, information unique to $X_1$, $\mu(X_1;Y)$, information unique to $X_2,$ $\mu(X_2;Y)$, and synergistic information, $\mu(X_1,X_2;Y)$, for some canonical examples from the literature \cite{prokopenko_quantifying_2014, griffith_quantifying_2015, kolchinsky_novel_2022, james_multivariate_2017}. The distribution of the examples considered are shown in Figure \ref{bivariate examples} together with the descriptors that minimize Equation (\ref{Icap}). Tthe two bottom lines show the values of shared information for two special conditions considered in \cite{kolchinsky_novel_2022}. The second condition corresponds to the identity property}
\label{Tab:two sources examples}
\end{table}

To analyze the geometrical properties of the examples in Table \ref{Tab:two sources examples} and their decomposition we introduce the following graphical convention to visualize the probability distributions $P(X, Y)$ for $X = \{ X_1, X_2 \}$, Figure \ref{bivariate examples}.
We denote with shapes $\{ \bm{\Circle}, \bm{\CIRCLE}, \bm{\times}, \bm{+}, \bm{\mdblksquare}, \dots \}$ the elements of $\mathbb{Y}$. We denote with points in $\mathbb{R}^2$ the elements of $\mathbb{X}$.
At each position $x \in \mathbb{Y}$ we draw the shape of those elements $y \in \mathbb{Y}$ for which $p(x, y) > 0$.
Next to each shape we indicate the value $p(x, y)$. For example, $\bm{\Circle}^{0.3}$ indicates $p(x, \bm{\Circle}) = 0.3$. 
If $P(X, Y)$ is constant for all $x \in X$ and all $y \in Y$, we omit the indication of the probability value, for convenience.
If $x=[x_1,x_2], x' = [x'_1, x'_2] \in \mathbb{X}$ are such that $x_n = x_n'$, for $n=1$ or $n=2$, we highlight this by connecting the two points in $\mathbb{R}^2$ with a line.

Distribution \textsc{Rdn}, Figure \ref{bivariate examples}\emph{a}, is the archetype of redundant information \cite{prokopenko_quantifying_2014}. Both $X_1$ and $X_2$ are isomorphic to $X$ and thus convey the same information about $Y$ than $X$ itself. Our decomposition returns one bit of information shared by $X_1$ and $X_2$. All other terms of the breakdown have zero information, Table \ref{Tab:two sources examples}.

We can transform the redundancy in \textsc{Rdn} into information unique to $X_1$ by imagining of moving the two points in \textsc{Rnd} so that they become aligned parallel to the first coordinate axis, Figure \ref{bivariate examples}\emph{b}. We call \textsc{Unq1} this distribution in which $X_1$ is still isomorphic to $X$ but all information available in $X$ is ``masked'' to $X_2$. Our decomposition returns one bit of information unique to $X_1$ and zero information for all other elements of the breakdown for this example, Table \ref{Tab:two sources examples}.
If we align the two points in \textsc{Rdn} parallel to the second coordinate axis, Figure \ref{bivariate examples}\emph{c}, we obtain the complementary distribution \textsc{Unq2} with one bit of information unique to $X_2$, Table \ref{Tab:two sources examples}.

We propose that \textsc{Unq1} and \textsc{Unq2} are the ``building blocks'' of unique information. Any distribution $P(X_1, X_2,Y)$ in which either $X_1$ and $X_2$ convey unique information must include some variation of the maskings between different outcomes of $Y$ observed in \textsc{Unq1} or \textsc{Unq2}.
For example, \textsc{Imperfect Rdn}, Figure \ref{bivariate examples}\emph{d}, can be thought of as ``leaking'' part of the redundant information \textsc{Rdn} into information unique to $X_2$ by means of masking some of the target information available in $X$ to $X_1$. Our decomposition provides 0.93\,bit of redundancy and 0.07\,bit of information unique to $X_2$, Table \ref{Tab:two sources examples}. Example \textsc{Unq}, Figure \ref{bivariate examples}e, is the archetype of unique information.
Using the descriptor previously shown in Figure \ref{Fig: Decompositions of UNQ}, our decomposition provides one bit of unique information both for $X_1$ and for $X_2$ and no redundant or synergistic information, Table \ref{Tab:two sources examples}.

We can generate a distribution with synergistic information by means of combining \textsc{Unq1} and \textsc{Unq2} to obtain example \textsc{Syn} shown in Figure \ref{bivariate examples}\emph{f}. An alternative way to derive \textsc{Syn} is to mirror either \textsc{Unq1} or \textsc{Unq2} along an imaginary axis with slope $\pi / 4$ (or $-\pi / 4$) and random intercept. For \textsc{Syn} we cannot devise a descriptor of $Y$ such that either $X_1$ or $X_2$ can convey the same information as $X$ for each singled-out features of $Y$. Our decomposition assigns half bit of synergistic information and half bit of redundant information, Table \ref{Tab:two sources examples}.
We propose that the two maskings involving the same two relizations of $Y$ in \textsc{Syn} are the ``building block'' of synergistic information.
A special case of \textsc{Syn} is the distribution \textsc{Corner} shown in Figure \ref{bivariate examples}\emph{g}. This distribution is of particular interest because it is found in all canonical examples of synergy discussed below. Our decomposition breaks the 0.92\,bit information of \textsc{Crossing} into 0.25\,bit of redundant information and 0.67\,bit of synergistic information, Table \ref{Tab:two sources examples}.

The \textsc{Xor} distribution, Figure \ref{bivariate examples}\emph{h}, is the archetype of synergistic information \cite{prokopenko_quantifying_2014} and can be thought of as the combination of four \textsc{Corner} elements. Accoding to our breakdown, all information conveyed jointly by $X_1$ and $X_2$ about $Y$ is synergistic, Table \ref{Tab:two sources examples}, as expected.

\input{figures/Fig_Two_sources_examples.tex}

Two further classic example of synergistic distribution are \textsc{And} and \textsc{Sum} \cite{prokopenko_quantifying_2014}, Figures \ref{bivariate examples}\emph{i} and \ref{bivariate examples}\emph{j}. Equation (\ref{muY}) breaks the 0.81\,bit of total information of \textsc{And} into 0.5\,bit of synergistic and approximately 0.31\,bits of redundant information, Table \ref{Tab:two sources examples}. For \textsc{Sum}, our decomposition breaks the 1.5\,bit of total information into 1\,bit of synergistic and 0.5\,bits of redundant information, Table \ref{Tab:two sources examples}.

Our decomposition also allocates the three bits of information in examples \textsc{Dyadic} and \textsc{Triadic} in a way that reflects the different generative structures of the two systems \cite{james_multivariate_2017}. Example \textsc{Dyadic}, Figure \ref{bivariate examples}\emph{k}, consists of two identical \textsc{Unq} structures \cite{james_multivariate_2017}. Our decomposition returns one bit of information for both unique terms, Table \ref{Tab:two sources examples}. Examples \textsc{Triadic} and \textsc{RdnXor}, Figures \ref{bivariate examples}\emph{l} and \ref{bivariate examples}\emph{m}, both consist of two \textsc{Xor}-like structures \cite{james_multivariate_2017}. Our decomposition breaks the two bits of information of these examples into one bit of synergystic and one bit of redundant information, Table \ref{Tab:two sources examples}.

For \textsc{RdnUNQXor} \cite{prokopenko_quantifying_2014}, Figure \ref{bivariate examples}\emph{n}, our decomposition provides one bit of informaiton for each term of the breakdown, Table \ref{Tab:two sources examples}, as expected.

Finally, for a proof that Equation (\ref{Icap}) satisfies $I(X_1 {:} X_2; X_1) = I(X_1; X_2)$ and $I(X_1 {:} X_2 ; \{ X_1, X_2 \}) = I(X_1;X_2)$, see the proof of the identity property, Section \ref{app:2sources:identity property}. 

%

\subsection{Properties of Equation (\ref{Icap})} \label{app:2sources:identity property}

\begin{theorem}[Williams and Beer axioms]
Equation (\ref{Icap}) satisfies the following properties:
\begin{enumerate}
\item
\emph{Symmetry}: $I(A_1 {:} \dots {:} A_K; Y)$ does not depend on the order of $A_1, \dots, A_K$
\item
\emph{Self redundancy}: $I(A {:} A; Y) = I(A;Y)$
\item
\emph{Monotonicity}: $I(A_1 {:} \dots {:} {A_K}; Y) \le I(A_1 {:} \dots {:} {A_{K-1}}; Y)$
\end{enumerate}
\end{theorem}
\begin{proof}
Property one follows from the the fact that the minimum operator $\min \{ a_1, \dots, a_K \}$ does not depend on the order of the elements of $\{ a_1, \dots, a_K \}$. Property two follows from the fact that equation $I(A {:} A; Y)$ reduces to decomposition (\ref{MI expansion}) for any choice of descriptor of $Y$. We define
$$
\mathcal{Y}' \triangleq \argmin_{\mathcal{Y} \in \Omega_Y}
\Big \{
I \big(A_1{:}\dots{:} A_K; \mathcal{Y})
\Big \}
\quad
\text{and}
\quad
\mathcal{Y}'' \triangleq \argmin_{\mathcal{Y} \in \Omega_Y}
\Big \{
I \big(A_1{:}\dots{:} A_{K-1}; \mathcal{Y})
\Big \}.
$$
Because $\min \{a_1, \dots, a_K \} \le \min \{a_1, \dots, a_{K-1} \}$ for any set $\{a_1, \dots, a_K\}$, we have
$$
I \big(A_1{:}\dots{:} A_K; Y) = I \big(A_1{:}\dots{:} A_K; \mathcal{Y}') \le I \big(A_1{:}\dots{:} A_K; \mathcal{Y}'') \le I \big(A_1{:}\dots{:} A_{K-1}; \mathcal{Y}'') = I \big(A_1{:}\dots{:} A_{K-1}; Y)
$$
This proves property three.
\end{proof}

\begin{theorem}[Identity property]
Equation (\ref{Icap}) satisfies $I(X_1 {:} X_2; \{X_1,X_2\}) = I(X_1; X_2)$.
\end{theorem}
\begin{proof}
In general, we have
\begin{align}
I(X_1 {:} X_2; Y) &
= \min_{\mathcal{Y} \in \Omega_Y}
\Big \{
I \big(X_1 {:} X_2; \mathcal{Y})
\Big \} 
=
\min_{\mathcal{Y} \in \Omega_Y}
\Bigg \{
\sum_{\ell=1}^L
\
\sum_{y^\ell }
p \big(y^\ell\big) \cdot \min_{k=1,2} \left \{ I\big(A_k;Y^{\ell-1} \mid y^\ell\big) \right \}
\Bigg \} = \nonumber \\
& \stackrel{(a)}{=}
\min_{\mathcal{Y} \in \Omega_Y}
\Bigg \{
I_1(X_1;\mathcal{Y}) + I(X_2;\mathcal{Y}) - 
\sum_{\ell=1}^L
\
\sum_{y^\ell }
p \big(y^\ell\big) \cdot \max_{k=1,2} \left \{ I\big(A_k;Y^{\ell-1} \mid y^\ell\big) \right \}
\Bigg \} = \nonumber \\
& \stackrel{(b)}{=}
I(X_1;Y) + I(X_2;Y) + 
\min_{\mathcal{Y} \in \Omega_Y}
\Bigg \{
-
\sum_{\ell=1}^L
\
\sum_{y^\ell }
p \big(y^\ell\big) \cdot \max_{k=1,2} \left \{ I\big(X_k;Y^{\ell-1} \mid y^\ell\big) \right \}
\Bigg \} = \nonumber \\
& \stackrel{(c)}{=}
I(X_1;Y) + I(X_2;Y) - 
\max_{\mathcal{Y} \in \Omega_Y}
\Bigg \{
\sum_{\ell=1}^L
\
\sum_{y^\ell }
p \big(y^\ell\big) \cdot \max_{k=1,2} \left \{ I\big(X_k;Y^{\ell-1} \mid y^\ell\big) \right \}
\Bigg \} = \nonumber \\
& \stackrel{(d)}{=}
I(X_1;Y) + I(X_2;Y) - U(X_1 {:} X_2; Y) \label{identity property last step} 
\end{align}
where (a) follows from the maximum-minimums identity, (b) from the fact that $I(X_1;Y)$ and $I(X_2;Y)$ do not depend on the choice of the descriptor, (c) from the fact that mutual information is non-negative, and (d) from the definition of union information, Equation (\ref{Icup}).

By construction $U(X_1 {:} X_2; Y) \le I(X_1,X_2;Y)$.
Furthermore, $U(X_1 {:} X_2; \mathcal{C}_Y) = I(X_1,X_2;Y) \le U(X_1 {:} X_2; Y) \le I(X_1,X_2;Y)$.
Substituting $U(X_1 {:} X_2; Y) = I(X_1,X_2;Y)$ in equation (\ref{identity property last step}) provides $I(X_1 {:} X_2; Y) = I(X_1;Y) + I(X_2;Y) - I(X_1,X_2;Y)$.
For $Y = \{X_1,X_2\}$ we have $I(X_1;Y) + I(X_2;Y) - I(X_1,X_2;Y) = I(X_1;X_2)$ . The theorem is proved.
\end{proof}

The same strategy can also be used to show that $I(X_1 {:} X_2; X_1) = I(X_1; X_2)$.

\begin{theorem}[Additivity property]
Assume that $\{ A_1, A_2, Y \}$ is independent of $\{ \hat{A}_1, \hat{A}_2, \hat{Y} \}$.
Equation (\ref{Icap}) satisfies
$$
I(\{ A_1, \hat{A}_1 \} {:} \{ A_2, \hat{A}_2 \}; \{ Y, \hat{Y} \} ) = I(A_1 {:} A_2; Y) + I(\hat{A}_1 {:} \hat{A}_2; \hat{Y})
$$
\end{theorem}

\begin{proof}
From Equation (\ref{identity property last step}) we have $I(A_1 {:} A_2; Y) = I(A_1;Y) + I(A_2; Y) - I(A_1, A_2;Y)$ and $I(\hat{A}_1 {:} \hat{A}_2; \hat{Y}) = I(\hat{A}_1;\hat{Y}) + I(\hat{A}_2; \hat{Y}) - I(\hat{A}_1, \hat{A}_2;\hat{Y})$, and also
\begin{align*}
I(\{ A_1, \hat{A}_1 \} {:} \{ A_2, \hat{A}_2 \}; \{ Y, \hat{Y} \} )
&=
I(A_1, \hat{A}_1; Y, \hat{Y}) + I(A_2, \hat{A}_2; Y, \hat{Y}) - I(A_1, \hat{A}_1, A_2, \hat{A}_2; Y, \hat{Y}) \\
&=
I(A_1; Y) + I(\hat{A}_1; \hat{Y}) + I(A_2; Y) + I(\hat{A}_2; \hat{Y}) - I(A_1, A_2; Y) - I(\hat{A}_1, \hat{A}_2; \hat{Y}).
\end{align*}
The theorem is proved.
\end{proof}

\begin{theorem}[Blackwell property]
Equation (\ref{Icap}) satisfies $I(X_1 {:} \dots {:} X_N; Y) = I(X_1 {:} \dots {:} X_{N-1};Y)$ if $X_N = f(X_n)$ for some $n \in \{ 1, \dots, N \}$.
\end{theorem}
\begin{proof}
The property follows from the data processing inequality.
\end{proof}

\begin{theorem}[Combined secret sharing property]
Equation (\ref{Icap}) satisfies the combined secret sharing property
$$
I(A_1 {:} \dots {:} A_K; S_1, \dots, S_L) = H(\{ S_\ell : A_1, \dots, A_K \in \mathcal{A}_i \})
$$
where $\mathcal{A}_1, \dots, \mathcal{A}_K$ are access structures\footnote{Refer to \cite{rauh_secret_2017} for a definition of the relevant quantities and notations.} of a combination of $L$ perfect secret sharing schemes.
\end{theorem}

\begin{proof}
The probabilistic independence of the secrets \cite{rauh_secret_2017} implies
$$
H(\{ S_\ell : A_1, \dots, A_K \in \mathcal{A}_i \}) ) =
\sum_{\{ S_\ell:A_1, \dots, A_K \in \mathcal{A}_i \}} H(S_\ell).
$$
Denote $S = \{ S_\ell:A_1, \dots, A_K \in \mathcal{A}_i \}$, we have
\begin{align}
I(A_1 {:} \dots {:} A_K; S_1, \dots, S_L) &
\stackrel{(a)}{\ge}
I(A_1 {:} \dots {:} A_K {:} S; S_1, \dots, S_L) \ge \\
& \stackrel{(b)}{\ge}
I(S; S_1, \dots, S_L) = H(S)
\end{align}
where (a) follows from the monotononicity property and (b) from the Blackwell property, given that $I(A_k; S) = H(S)$ for $k = 1 \dots, K$, implies $f_k(A_k) = S$ exists for each $k$.
For the canonical descriptor $\mathcal{C}_{\{ S_1, \dots, S_L \}} = Y^0 \rightarrow \dots \rightarrow Y^L$ we have  $I(A_k;Y^{\ell-1} \mid Y^\ell) = I(A_k; S_\ell \mid S_{\ell+1}, \dots, S_L) = I(A_k; S_\ell)$. Thus $min_{k=1,\dots,K} \big \{ I(A_k; Y^{\ell-1} \vert y^\ell) \big \} = min_{k=1,\dots,K} \big \{ I(A_k; Y^{\ell-1}) \big \}$, which is equal to $H(S_\ell)$ if $A_1, \dots, A_K \in \mathcal{A}_L$, and is zero otherwise. This also implies that $I(A_1 {:} \dots {:} A_K; \mathcal{Y}) = H(S)$. The theorem is proved.
\end{proof}

\subsection{Proof of the properties of Equation (\ref{multipleMI})} \label{app:proof property 4}

\begin{theorem}[Non-negativity]
Equation (\ref{multipleMI}) satisfies $I(X_1; \dots; X_N) \ge 0$.
\end{theorem}
\begin{proof}
The property follows from the non-negativity of Equation (\ref{Icap}).
\end{proof}

\begin{theorem}[Symmetry]
Equation (\ref{multipleMI}) is invariant to permutations of $X_1, \dots, X_N$.
\end{theorem}
\begin{proof}
The property follows from the symmetry of Equation (\ref{Icap}).
\end{proof}

\begin{theorem}[Monotonicity]
Equation (\ref{multipleMI}) satisfies $I(X_1; \dots; X_{N-1}) \ge I(X_1; \dots; X_N)$ for all $P(X_1, \dots, X_N)$.
\end{theorem}
\begin{proof}
We have $I(X_1; \dots; X_N) = I(X_1 {:} \dots {:} X_N; X_1, \dots, X_N) \le I(X_1 {:} \dots {:} X_{N-1}; X_1, \dots, X_N)$ from the monotonicity of Equation (\ref{Icap}). We note that, in general, $\Omega_{\{X_1, \dots, X_{N-1}\}} \subseteq \Omega_{\{X_1, \dots, X_N\}}$. The minimization in Equation (\ref{Icap}) thus guarantees that $I(X_1 {:} \dots {:} X_{N-1}; X_1, \dots, X_N) \le I(X_1 {:} \dots {:} X_{N-1}; X_1, \dots, X_{N-1}) = I(X_1; \dots; X_{N-1})$.
\end{proof}

\begin{theorem} [Blackwell property]
Equation (\ref{multipleMI}) satisfies the following properties
\begin{enumerate}
\item
$I(X_1; \dots; X_N) = I(X_1; \dots; X_{N-1})$ if $X_{N-1} = f(X_N)$ for some $f:X_N \rightarrow X_{N-1}$;
\item
$I(X_1; \dots; X_N) = H(X_1)$ if $X_1, \dots, X_N$ form a Markov chain $X_N \rightarrow X_{N-1} \rightarrow \dots \rightarrow X_1$
\end{enumerate}
\end{theorem}

\begin{proof}
Property one follows from Equation (\ref{Icap}) and from the signal processing inequality. Property two follows directly from property one.
\end{proof}

\begin{theorem} [Shannon property]
Equation (\ref{multipleMI}) satisfies the following properties
\begin{enumerate}
\item
$\exists \bar{n}\in \{ 1, \dots, N \}: p(x_{\bar{n}}, x_n) = p(x_{\bar{n}}) \cdot p(x_n)$ for all $n \ne \bar{n}$ and all $x \in X \implies I(X_1; \dots; X_N) = 0 $.
\item
$p(x_1, \dots, x_N) = p(x_1) \cdot \dots \cdot p(x_N)$ for all $x \in X \implies I(X_1; \dots; X_N) = 0$.
\item
$I(X_1; \dots; X_N) = 0 \centernot\implies p(x_{n_1}, x_{n_2}) = p(x_{n_1}) \cdot p(x_{n_2})$ for all $n_1, n_2 = 1, \dots, N$
\end{enumerate}
\end{theorem}

\begin{proof}
To prove property one, without loss of generality thanks to the symmetry of Equation (\ref{multipleMI}), we assume $\bar{n} = N$. We consider $Y = X$. For the canonical descriptor $\mathcal{C}_Y = Y^0 \rightarrow \dots, \rightarrow Y^N$ we obtain
$$
min_{n=1,\dots,N} \big \{ I(X_n;Y^{n-1} \mid y^n) \big \} = 0,
\quad
n=1,\dots,N-1,
\quad
\forall y^n \in Y^n
$$
Furthermore
$$
min_{n=1,\dots,N} \big \{ I(X_n; Y^{N-1} \mid y^N) \big \} =
min_{n=1,\dots,N} \big \{ I(X_n; Y^{N-1}) \big \} =
min_{n=1,\dots,N} \big \{ I(X_n; X_N) \big \} = I(X_n; X_{\bar{n}}) = 0
$$
This proves the first property. Property two is follows directly from property one.

To prove property three consider the counterexample in Table \ref{Tab:counterexample} for the case $N=3$. We obtain $I(X_2; Y^0 \mid Y^1{=}0) = I(X_3; Y^0 \mid Y^1{=}1) = I(X_1; Y^1 \mid Y^2) = 0$, thus $I(X_1; X_2; X_3) = 0$. However $I(X_1;X_2) = 1$, $I(X_1;X_3) = 0.5$, $I(X_2;X_3) = 1$.

\begin{table}[t!]
\centering
\begin{tabular}[t]{ | c c c c | c c c |}
	\hline
	\rule{0pt}{11pt}$X_1$ 	& $X_2$ 	& $X_3$ 	& $P(X,Y)$				&	$Y^0$	&	$Y^1$	&	$Y^2$	\\
	\hline
	0 		& 0 		& 1 		& $\nicefrac{1}{4}$		&	0		&	0		&	0		\\
	1 		& 0 		& -1 		& $\nicefrac{1}{4}$		&	1		&	0		&	0		\\
	0 		& 1 		& 0 		& $\nicefrac{1}{4}$		&	2		&	1		&	0		\\
	1 		& -1 		& 0 		& $\nicefrac{1}{4}$		&	3		&	1		&	0		\\
	\hline
\end{tabular}
\caption{Counterxample demonstrating property three of the Shannon property for the case $N=3$.}
\label{Tab:counterexample}
\end{table}
\end{proof}

\end{document}